\documentclass[twocolumn, twocolappendix]{aastex63}

\shorttitle{Measuring Hot Blackbodies}
\shortauthors{Arcavi}

\graphicspath{{./}}

\begin{document}

\title{Errors When Constraining Hot Blackbody Parameters with Optical Photometry}

\author[0000-0001-7090-4898]{Iair Arcavi}
\affiliation{The School of Physics and Astronomy, Tel Aviv University, Tel Aviv, 69978, Israel}
\affiliation{CIFAR Azrieli Global Scholars program, CIFAR, Toronto, Canada}

\correspondingauthor{Iair Arcavi}
\email{arcavi@tauex.tau.ac.il}

\begin{abstract}
Measuring blackbody parameters for objects hotter than a few $10^4$\,K  with optical data alone is common in many astrophysical studies. However this process is prone to large errors because at those temperatures the optical bands are mostly sampling the Rayleigh–Jeans tail of the spectrum. Here we quantify these errors by simulating different blackbodies, sampling them in various bands with realistic measurement errors, and refitting them to blackbodies using two different methods and two different priors. We find that when using only optical data, log-uniform priors perform better than uniform priors. Still, measured temperatures of blackbodies above $\sim$35,000\,K can be wrong by $\sim$10,000\,K, and only lower limits can be obtained for temperatures of blackbodies hotter than $\sim$50,000\,K. Bolometric luminosities estimated from optical-only blackbody fits can be wrong by factors of 3--5. When adding space-based ultraviolet data, these errors shrink significantly. For when such data are not available, we provide plots and tables of the distributions of true temperatures that can result in various measured temperatures. It is important to take these distributions into account as systematic uncertainties when fitting hot blackbodies with optical data alone. 
\end{abstract}

\keywords{High energy astrophysics (739), Astronomical methods (1043), Optical astronomy (1776), Ultraviolet astronomy (1736)}

\section{Introduction}

In many astrophysical studies it useful to fit a blackbody spectrum to broadband photometry in order to determine the effective temperature, emitting radius, and bolometric luminosity of an object or transient event. Hot blackbodies (at temperatures of a few $10^4$\,K) are of particular interest, for example in certain tidal disruption events (see \citealt{vanVelzen2020} and \citealt{Gezari2021} for recent reviews), in the early phases of supernovae \citep[e.g.][]{Valenti2016} and in the very early phases of kilonovae \citep[at least based on the one case observed so far;][]{Abbott2017}.

Constraining the temperature, and with it the bolometric luminosity, of tidal disruption events is important for distinguishing between emission models, and for deriving properties of the supermassive black hole population \citep[e.g.][]{Piran2015,Dai2018,Mockler2019,Ryu2020}. Measuring the cooling rate of supernovae during their first days can be used to constrain both the progenitor parameters and explosion physics \citep[e.g.][]{Nakar2010,Rabinak2011,Shussman2016,Arcavi2017,Rubin2017,Sapir2017}. For kilonovae to be used to constrain the neutron-star equation of state, heavy-element nucleosynthesis, and even cosmology, it turns out that measuring their cooling during the first few hours is especially important \citep[e.g.][]{Arcavi2018}.

\begin{figure}[t]
\includegraphics[width=\columnwidth]{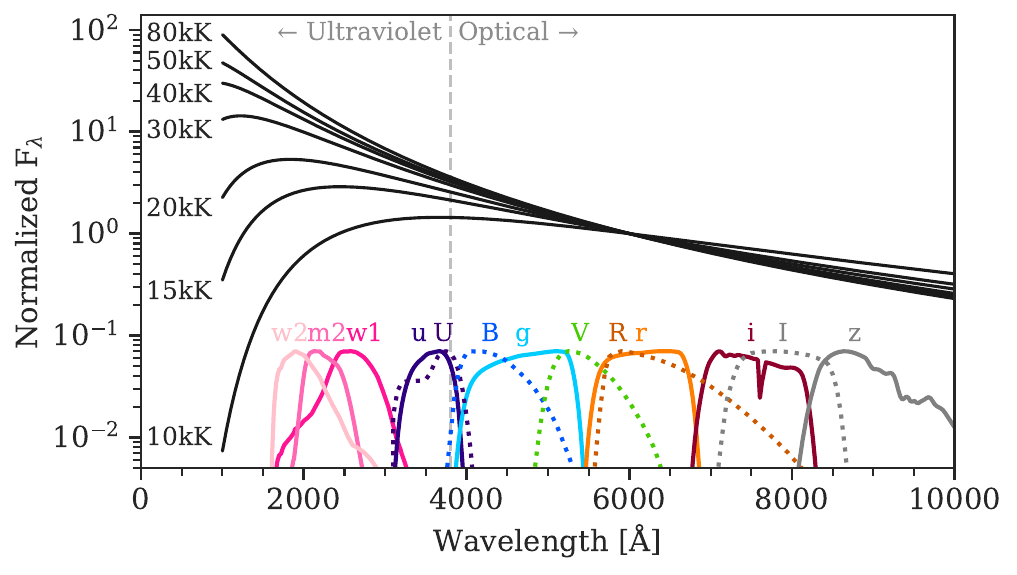}
\caption{\label{fig:filters}Blackbody spectra at various temperatures (denoted in kilo-Kelvin), normalized to their flux at 6,000\,$\textrm{\AA}$. {\it Swift} $uvw1$ (abbreviated as $w1$), $uvm2$ ($m2$) and $uvw2$ ($w2$) filter response curves are shown at the bottom together with those of the optical-infrared $ugriz$ (solid lines) and $UBVRI$ (dotted lines) bands. Blackbodies at temperatures $\gtrsim$30,000\,K all appear very similar in the optical to near-infrared regime.} 
\end{figure}

However, in many of these cases (when not highly redshifted), the optical wavelength regime is in the Rayleigh–Jeans tail of the spectrum, where it is difficult to constrain the temperature with optical data alone. Figure \ref{fig:filters} illustrates how the spectra of blackbodies hotter than approximately 30,000\,K are very similar in the optical bands. It is therefore expected that optical data alone will not be able to distinguish between blackbodies at those temperatures clearly, especially when measurement uncertainties are taken into account.

Indeed, \cite{Faran2018} show that when approaching temperatures of 20,000\,K, temperature estimates based on optical data alone can be off by $\gtrsim$1,000\,K. Here we expand the quantification of such measurement errors and characterize them also at higher temperatures. We simulate observations of blackbodies at various temperatures, taking into account realistic measurement uncertainties, in both optical and ultraviolet bands, and fit the data to blackbodies as would be done for real observations (Section \ref{sec:method}). We then measure how different the best-fit temperature is from the true simulated temperature for different temperatures, band combinations, methods, priors, and measurement uncertainties (Section \ref{sec:results}). These results are then inverted in order to determine, given an optical-only temperature measurement, what true temperature values could produce it. We also check what effect the temperature measurement errors have on the errors of the deduced bolometric luminosities (Section \ref{sec:analysis}). 

We do not consider real-life complications such as K-corrections, distance uncertainties, extinction uncertainties, line emission and absorption contamination, and line blanketing. Extinction and line blanketing are especially important to consider with ultraviolet data; however here we wish to isolate the degeneracies and systematic errors of fitting just a hot blackbody component with optical data, and compare it to using also ultraviolet data.

\section{Method}\label{sec:method}

We use two separate methods for generating synthetic magnitudes from blackbody spectra and for fitting them back to blackbody spectra (each method performs both operations). The first method makes use of the Astrolib PySynphot package (hereafter referred to only as \texttt{pysynphot}; \citealt{pysynphot}), and the second uses the Light Curve Fitting package (hereafter, \texttt{lightcurve\_fitting}; \citealt{lightcurve_fitting}), version 0.4.0. 

The main difference between the methods is in the way the blackbody spectrum is fit to the data (see below). However, there are also differences in the magnitudes generated by each method for given blackbodies (Appendix \ref{sec:app-diffs} and Figure \ref{fig:methodsdiff}). Given these differences, we use each method to fit only data simulated with that same method. 

\begin{figure}
\includegraphics[width=\columnwidth]{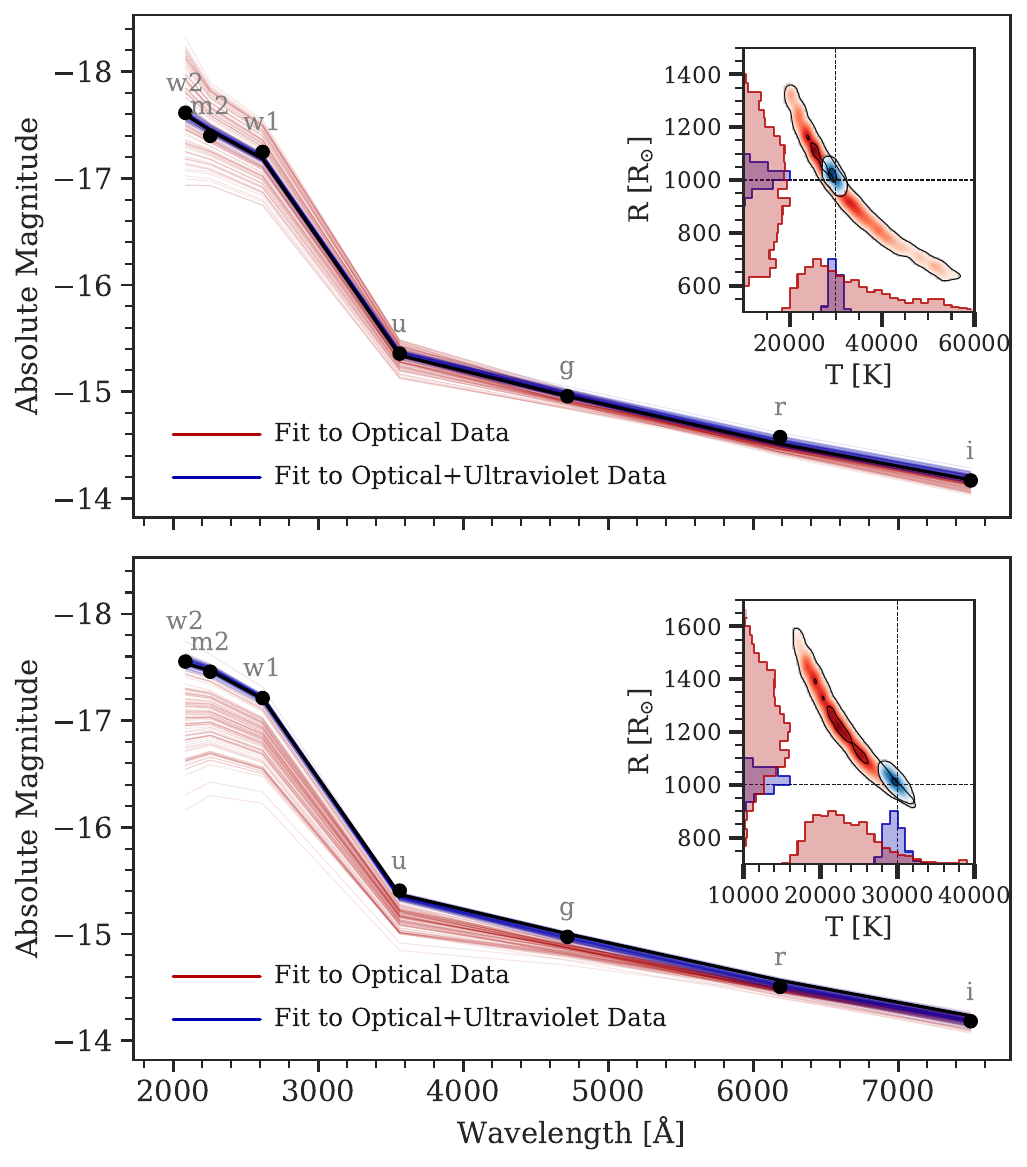}
\caption{\label{fig:eg_fits}Example MCMC fits to simulated magnitudes of a 30,000\,K, 1,000\,R$_{\sun}$ blackbody (black circles) assuming 0.05-magnitude photometric uncertainties, using only optical (in red) vs. optical+ultraviolet (in blue) bands (100 random models drawn from the posterior distribution are shown). The insets show the corner plots for each MCMC fit, with contour lines denoting 90\%  bounds, and dotted lines denoting the true values. The top and bottom plots are two different realizations of simulated data, each fit also with a different method (see text for details). Including ultraviolet data reduces both the errors and the uncertainties significantly.} 
\end{figure}

\begin{figure*}[t]
\includegraphics[width=\textwidth]{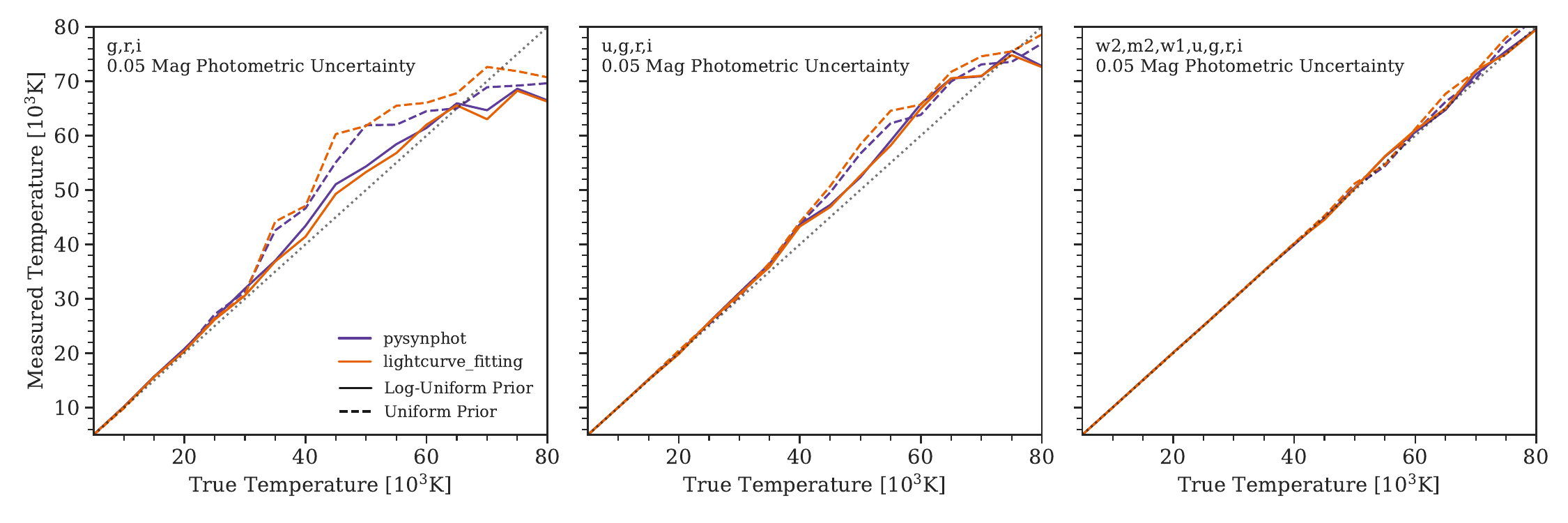}
\caption{\label{fig:res_t_sdss_p05_priorcomp}Median (of 100-realization ensembles) best-fit measured temperature vs. the true (simulated) temperature for different methods, priors, and band combinations as noted, using photometric uncertainties of 0.05 magnitudes. The fits using log-uniform priors (solid lines) produce more precise median temperatures compared to those using uniform priors (dashed lines), when using optical data alone (left panel). When adding space-based ultraviolet data (right panel), both priors perform equally well.}
\end{figure*}

\begin{figure*}[t]
\includegraphics[width=\textwidth]{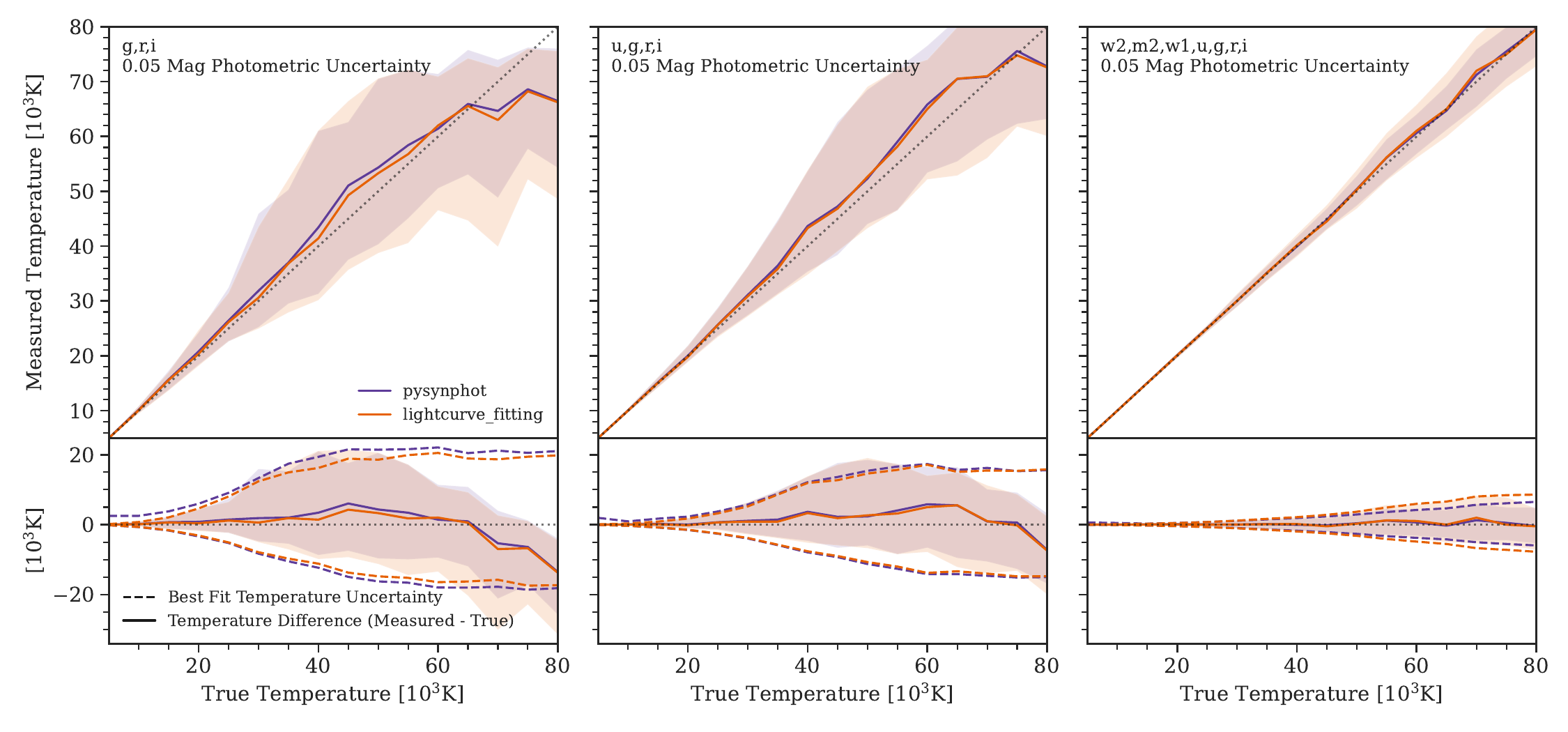}
\caption{\label{fig:res_t_sdss_p05}Top panels: best-fit measured temperature vs. the true (simulated) temperature for different band combinations as noted, using simulated measurement uncertainties of 0.05 magnitudes and log-uniform priors. The solid lines denote the ensemble median best-fit temperature, and the shaded regions denote the 16th--84th ensemble percentile. Results for both methods are shown. Bottom panels: same as top, but for the difference between the measured and true temperatures. The mean ensemble upper and lower fit uncertainties are shown in dotted lines.}
\end{figure*}

\begin{figure*}
\includegraphics[width=\textwidth]{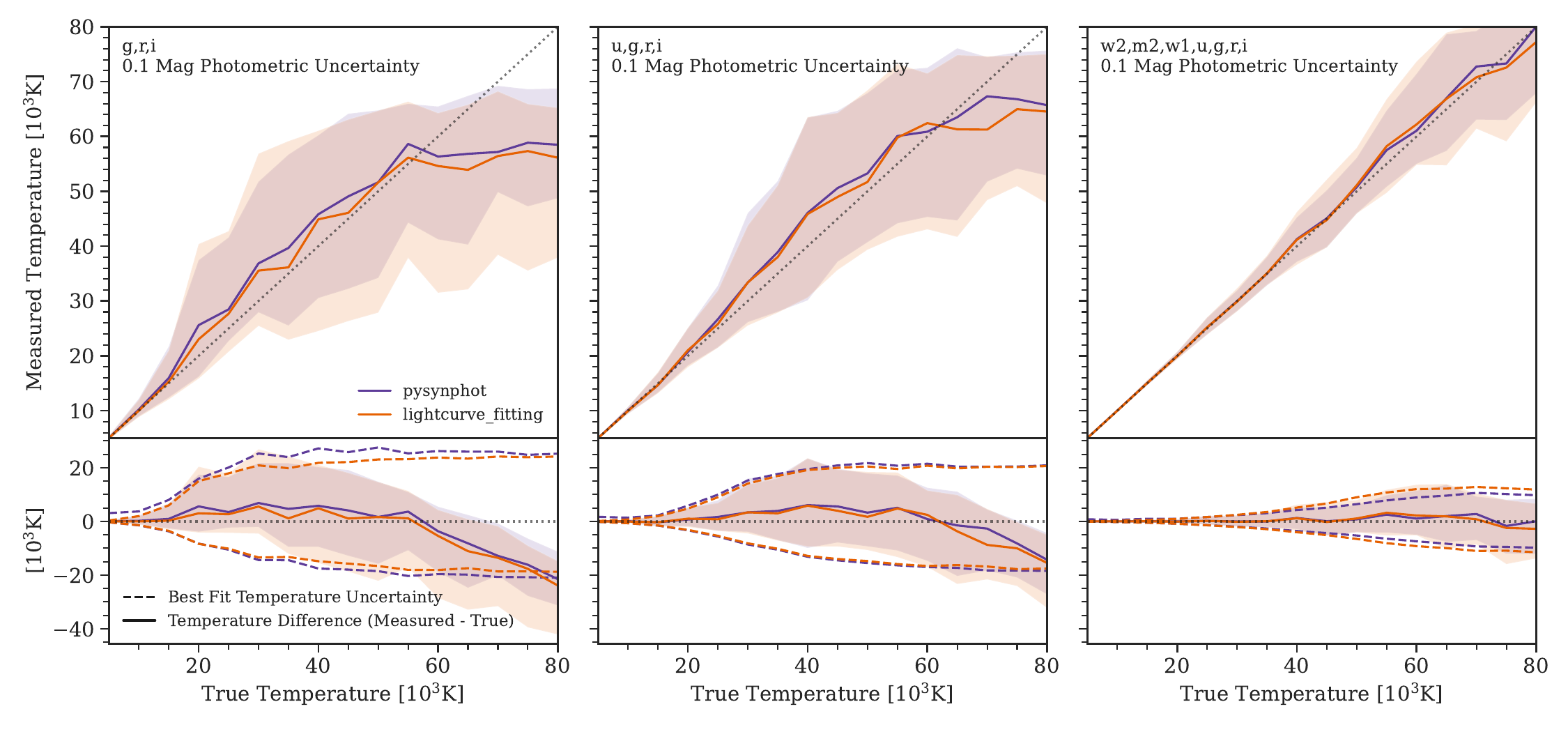}
\caption{\label{fig:res_t_sdss_p1}Same as Figure \ref{fig:res_t_sdss_p05} but using simulated photometric uncertainties of 0.1 magnitudes.}
\end{figure*}

\subsection{Generating Synthetic Data}

We create spectra of blackbodies with a radius of 1,000\,R$_{\sun}$ and temperatures of 5,000\,K--80,000\,K in 5,000\,K increments. For each blackbody, we generate synthetic spectral energy distribution (SED) magnitudes in the Sloan Digital Sky Survey \citep[SDSS;][]{Doi2010} $ugri$ bands, the Johnson-Cousins \citep{Johnson1953,Cousins1976} $BVRI$ bands, and The Neil Gehrels {\it Swift} Observatory (hereafter {\it Swift}; \citealt{Gehrels2004}) Ultraviolet/Optical Telescope \citep{Roming2005} $uvw1$, $uvm2$ and $uvw2$ bands (hereafter $w1$, $m2$ and $w2$ respectively)\footnote{Magnitudes in the $ugri$-bands are generated in the AB system, while magnitudes in the rest of the bands are generated in the Vega system.}. 

We generate data in six band combinations: $gri$ (optical observations), $ugri$ (optical with ground-based ultraviolet observations), $ugri$ + $w1$, $m2$ and $w2$ (optical with ground- and space-based ultraviolet observations), and similarly for the Johnson-Cousins bands, $BVRI$, $UBVRI$, and $UBVRI$ + $w1$, $m2$ and $w2$. We do not consider infrared bands. Although hot blackbodies differ in their infrared emission (Fig. \ref{fig:filters}), in practice their infrared emission is faint compared to their optical emission, and therefore more difficult to measure.

In total we produce 32 sets of blackbody simulation parameters (16 temperatures using two methods for generating the magnitudes).  For each of these 32 sets of parameters, we generate two ensembles of 100 SED realizations each by adding randomly generated Gaussian noise to each synthetic magnitude, once with a standard deviation of 0.05 magnitudes and once with a standard deviation of 0.1 magnitudes\footnote{We use fixed standard deviations for all bands at all temperatures, though in reality, measurement uncertainties could depend on band and temperature.}. 

\subsection{Fitting Blackbodies to the Synthetic Data}

We use the ``forward modeling'' technique preferred by \cite{Brown2016}, whereby the model spectrum (in our case, a blackbody) is convolved with the various filter response curves and then compared to the simulated data with chi-squared used as the likelihood function and all bands weighted equally. In the \texttt{pysynphot} method, the fit is done in magnitude vs. wavelength space, while in the \texttt{lightcurve\_fitting} package it is done in flux density ($F_\nu$) vs. frequency space. 

In both cases the fit is performed using the Markov Chain Monte Carlo (MCMC) technique as implemented by the \texttt{emcee} package \citep{Foreman-Mackey2013}. We use 16 walkers and 600 steps (of which 200 are burn-in steps) to fit both the temperature and radius of the blackbody to the data. We repeat the fits with two types of priors: uniform and log-uniform for both the temperature (in the range $10^3$--$10^5$\,K), and the radius (in the range $10$--$10^6$\,R$_{\sun}$). For the log-uniform case, we also test the effect of underestimating the photometric uncertainties by reporting half of the true uncertainty to the fitters. The code used to simulate and fit the blackbodies is publicly available on Github\footnote{\url{https://github.com/arcavi/bb_sims}}.

In total we fit each of the $32\times200$ realizations of simulated data in 18 different ways (six band combinations, using two types of priors and underestimated uncertainties for one of the priors). Thus we have 115,200 different fits, which we proceed to analyse.

\subsection{Nomenclature}

Hereafter, we will use the term ``error'' to denote the difference between a measured value and a true value, while the term ``uncertainty'' will denote the estimated spread of a measured value due to measurement errors or fit posteriors. For example, if for a 35,000\,K blackbody we measure a temperature of 29,000$\pm$7,000\,K, then the error is 6,000\,K while the uncertainty is 7,000\,K. 

\section{Results}\label{sec:results}

\subsection{Example Fits}

Example fits using the log-uniform priors for a 30,000\,K blackbody are shown in Figure \ref{fig:eg_fits}. As expected, fitting such a blackbody with optical data alone results in both inaccurate (i.e. having a large error) and imprecise (i.e. having a large uncertainty) measurements of the blackbody radius and temperature, compared to fitting it with optical and ultraviolet data together. In the top realization in Figure \ref{fig:eg_fits}, fit with the \texttt{pysynphot} method, the most likely values are close to the true ones, but the uncertainty in the optical-only fits are much larger compared to the fits which use the optical and ultraviolet data together. In the bottom realization, fit with the \texttt{lightcurve\_fitting} method, the optical-only fits show not only larger uncertainties compared to the combined optical-ultraviolet fits, but are also farther from the true values (by $\sim$10,000\,K). 

The difference in accuracy in this example is not due to the different methods (see below), but rather the different realizations. This is why we generate ensembles of 100 realizations for each temperature, band combination, magnitude uncertainty, and method, and look at the statistical properties of the resulting fits.

\subsection{Full Set of Fits}

For each individual MCMC fit we take the 50th percentile (i.e. median) of the posterior as the best-fit estimated value. Repeating this for each of the 100 realizations of a given ensemble produces an ensemble distribution of best-fit values for that set of blackbody temperature, band combination, magnitude uncertainty, and method. Similarly, we take the 16th and 84th percentiles of each MCMC fit posterior as the lower and upper uncertainty estimates of the fit respectively\footnote{We use the median and 68\% credible interval range since it is equivalent to using the mean $\pm1\sigma$ range for normally distributed results, but does not require assuming that the distribution is normal (or even symmetric).}, and also produce an ensemble-distribution of these values per parameter set.

\subsubsection{The Effect of Priors}

We plot the median of the best-fit temperature ensemble distributions for each set of blackbody simulation parameters and each type of prior for the 0.05 magnitude uncertainty simulations and the SDSS SEDs in Figure \ref{fig:res_t_sdss_p05_priorcomp}. When using optical and ultraviolet data (right panel in Figure \ref{fig:res_t_sdss_p05_priorcomp}), both priors produce similar results. However, when using optical data alone (left panel in Figure \ref{fig:res_t_sdss_p05_priorcomp}), the log-uniform priors produce more accurate ensemble median best-fit temperatures, compared to the uniform priors. The results for the SDSS 0.1 magnitude uncertainties and for both uncertainty values with the Johnson-Cousins SEDs are similar and are presented in Figures \ref{fig:res_t_sdss_p1_priorcomp}, \ref{fig:res_t_john_p05_priorcomp} and \ref{fig:res_t_john_p1_priorcomp}. We conclude that the log-uniform priors produce more precise results when fitting optical data, and we proceed to analyse only those fits in the remainder of this work. 

\subsubsection{Full Results for the Log-Uniform Priors}

We plot the median and 16th--84th percentile of the ensemble distributions for each set of blackbody simulation parameters in Figures \ref{fig:res_t_sdss_p05} and \ref{fig:res_t_sdss_p1} for the SDSS SEDs (the results for the Johnson-Cousins SEDs are similar and are presented in Figures \ref{fig:res_t_john_p05} and \ref{fig:res_t_john_p1}). 

So, for example, if the true temperature of the object being measured is 40,000\,K, then $gri$-band measurements with an uncertainty of 0.05 magnitudes yield best-fit temperatures in the range $\sim$30,000--60,000\,K (68\% bounds; top left panel of Figure \ref{fig:res_t_sdss_p05}). The fit uncertainties in this case are $\sim$15,000--20,000\,K (bottom left panel of Figure \ref{fig:res_t_sdss_p05}). Incorporating ground- and space-based ultraviolet measurements reduces the range of the best-fit temperatures to just a few hundred K around the true temperature (top right panel of Figure \ref{fig:res_t_sdss_p05}). The fit uncertainties are also (correctly) reduced to a few hundred K in this case (bottom right panel of Figure \ref{fig:res_t_sdss_p05}).

\subsubsection{The Effect of Underestimating the Photometric Uncertainties}

We plot the simulation results in the same way as above for the fits where the reported uncertainty is half of the true one in Figures \ref{fig:res_t_sdss_p05_err_und} -- \ref{fig:res_t_john_p1_err_und}. We find that there is no significant difference compared to the case where the uncertainties are estimated correctly. The only apparent difference is that underestimating the photometric uncertainties causes the \texttt{pysynphot} method to underestimate the derived temperature uncertainties.

\section{Analysis} \label{sec:analysis}

Having established that the log-uniform priors perform better compared to the uniform priors and that underestimating the photometric uncertainties does not have a strong effect on the results, our analysis hereafter focuses only on the fits using the log-uniform priors and having the correct uncertainty estimations.

\subsection{Errors in Temperature Estimation}

We find that for 0.05 magnitude uncertainties (Fig. \ref{fig:res_t_sdss_p05}), relying on $gri$-band data alone can result in several thousand K errors in temperature (in both directions) already at 20,000\,K. At around 40,000\,K these errors increase to -10,000\,K and +20,000\,K. At temperatures of 80,000\,K, systematic underestimates of up to 20,000\,K can occur. For 0.1 magnitude uncertainties (Fig. \ref{fig:res_t_sdss_p1}), at temperatures of 30,000\,K, overestimates of 20,000\,K can occur, while at 80,000\,K, underestimates reach 40,000\,K. Adding $u$-band observations slightly decreases these errors, but not significantly. 

In all of these cases, the reported fit uncertainties (dotted lines in the bottom panels of Figures \ref{fig:res_t_sdss_p05} and \ref{fig:res_t_sdss_p1}) roughly encompass the error correctly. However, for blackbodies hotter than about 50,000\,K, the fit uncertainties remain symmetric while the error is systematically toward underestimating the temperature (as expected when sampling a blackbody deep in its Rayleigh–Jeans tail). Similar results are seen for the Johnson-Cousins SEDs (Figures \ref{fig:res_t_john_p05} and \ref{fig:res_t_john_p1}).

It is only when adding data from the {\it Swift} bands do the errors decrease significantly, and the fit uncertainties encompass the errors correctly at all temperatures. 

In all cases studied there are almost no differences between the two fitting methods.

\subsection{The True Temperature Posterior For Optical Fits}

Unfortunately, ultraviolet observations are expensive and not always possible. Therefore it desirable to know the correct uncertainty when measuring temperatures with optical data alone. This is essentially the action of inverting the top left plots of Figures \ref{fig:res_t_sdss_p05}, \ref{fig:res_t_sdss_p1}, \ref{fig:res_t_john_p05} and \ref{fig:res_t_john_p1}.

In other words, assume we are only able to obtain $gri$-band data, we fit a blackbody to them, and we obtain a best-fit temperature of 30,000\,K - what true temperatures could have led to that measurement, given a known magnitude uncertainty?

\begin{figure}
\includegraphics[width=\columnwidth]{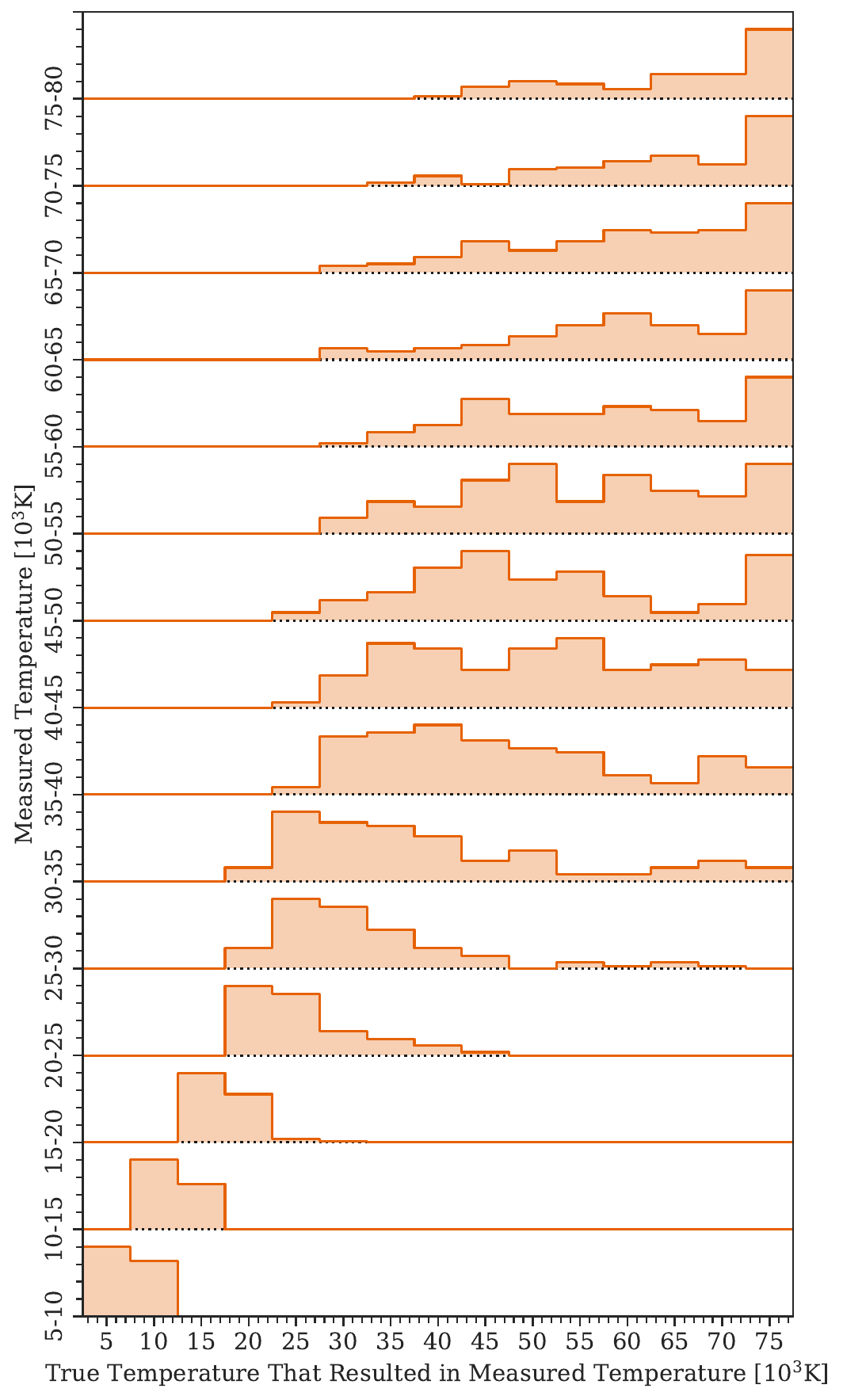}
\caption{\label{fig:inverse_t_eg}Distributions of the true temperatures (up to the limit of what was simulated - 80,000\,K) that lead to various measured temperatures (in 5,000\,K bins) when a blackbody is fit to $gri$-band data with 0.05 magnitude photometric uncertainties with the \texttt{lightcurve\_fitting} method. Above measured temperatures of 25,000\,K, the real temperature uncertainty becomes large.} 
\end{figure}

The answer to that question, for blackbodies fit to $gri$-band data with 0.05 magnitude uncertainties using the \texttt{lightcurve\_fitting} method, is presented in Figure \ref{fig:inverse_t_eg}. So, for example, if we obtained a best-fit temperature in the range 30,000--35,000\,K, the true temperature is indeed likely in that range, but could be as low as 20,000\,K and there is a probability tail going up to and above 60,000\,K (6th histogram from the bottom in Figure \ref{fig:inverse_t_eg}). The results are similar for $BVRI$-band data and when using the \texttt{pysynphot} method. These are shown, together with 0.1 magnitude uncertainty data, in Figure \ref{fig:inverse_t_all}. 

The 16th, 50th, and 84th percentiles of the true temperature distributions that produce a given measured temperature range are presented (as the median and lower and upper bounds) in Tables \ref{tab:inverse_t_flux} (for the \texttt{lightcurve\_fitting} method) and \ref{tab:inverse_t_mag} (for the \texttt{pysynphot} method). The data behind these distributions are publicly available on GitHub\footnote{\url{https://github.com/arcavi/bb_sims}}. Continuing our example where we obtained a best-fit temperature in the range 30,000--35,000\,K by fitting $gri$-band data with 0.05 magnitude uncertainties, the true temperature could be between 25,000\,K and 55,000\,K (68\% confidence bounds) with a median of 35,000\,K. 

This is a large uncertainty, which needs to be taken into account when fitting hot blackbodies using optical data alone. It becomes especially important for optical-only measured temperatures $\gtrsim$ 25,000\,K.

\begin{deluxetable*}{@{\extracolsep{4pt}}ccccc@{}}
\tablecaption{Median and 68\% confidence bounds of the true temperatures that produce various measured temperatures in the \texttt{lightcurve\_fitting} method for different band and magnitude uncertainty combinations.} \label{tab:inverse_t_flux}
\tablehead{
\colhead{Measured temperature} & \multicolumn{4}{c}{True temperature} \\
\cline{1-1} \cline{2-5}
\colhead{} & \multicolumn{2}{c}{For a $gri$-band measurement} & \multicolumn{2}{c}{For a $BVRI$-band measurement}\\
\cline{2-3} \cline{4-5}
\colhead{} & \colhead{0.05 mag uncertainty} & \colhead{0.1 mag uncertainty} & \colhead{0.05 mag uncertainty} & \colhead{0.1 mag uncertainty} \\
\colhead{($10^3$\,K)} & \colhead{($10^3$\,K)} & \colhead{($10^3$\,K)} & \colhead{($10^3$\,K)} & \colhead{($10^3$\,K)}}
\startdata
5--10 & $5.0_{ -0.0 }^{ +5.0 }$ & $5.0_{ -0.0 }^{ +5.0 }$ & $10.0_{ -5.0 }^{ +0.0 }$ & $5.0_{ -0.0 }^{ +5.0 }$ \\
10--15 & $10.0_{ -0.0 }^{ +5.0 }$ & $15.0_{ -5.0 }^{ +0.0 }$ & $15.0_{ -5.0 }^{ +0.0 }$ & $12.5_{ -2.5 }^{ +2.5 }$ \\
15--20 & $15.0_{ -0.0 }^{ +5.0 }$ & $20.0_{ -5.0 }^{ +15.0 }$ & $20.0_{ -5.0 }^{ +0.0 }$ & $20.0_{ -5.0 }^{ +5.0 }$ \\
20--25 & $25.0_{ -5.0 }^{ +5.0 }$ & $35.0_{ -15.0 }^{ +16.0 }$ & $25.0_{ -5.0 }^{ +0.0 }$ & $25.0_{ -5.0 }^{ +15.0 }$ \\
25--30 & $30.0_{ -5.0 }^{ +10.0 }$ & $35.0_{ -10.0 }^{ +29.2 }$ & $30.0_{ -5.0 }^{ +10.0 }$ & $35.0_{ -10.0 }^{ +20.0 }$ \\
30--35 & $35.0_{ -10.0 }^{ +20.0 }$ & $40.0_{ -15.0 }^{ +25.6 }$ & $35.0_{ -10.0 }^{ +10.0 }$ & $45.0_{ -18.2 }^{ +23.2 }$ \\
35--40 & $45.0_{ -10.0 }^{ +20.0 }$ & $50.0_{ -20.0 }^{ +20.0 }$ & $40.0_{ -10.0 }^{ +15.0 }$ & $45.0_{ -10.0 }^{ +20.0 }$ \\
40--45 & $50.0_{ -15.0 }^{ +20.0 }$ & $47.5_{ -20.3 }^{ +22.5 }$ & $45.0_{ -10.0 }^{ +25.0 }$ & $50.0_{ -15.4 }^{ +20.0 }$ \\
45--50 & $50.0_{ -10.0 }^{ +25.0 }$ & $55.0_{ -20.0 }^{ +20.0 }$ & $52.5_{ -12.5 }^{ +17.5 }$ & $50.0_{ -15.0 }^{ +20.0 }$ \\
50--55 & $55.0_{ -15.0 }^{ +15.2 }$ & $60.0_{ -25.0 }^{ +15.0 }$ & $50.0_{ -10.0 }^{ +20.0 }$ & $55.0_{ -20.0 }^{ +15.0 }$ \\
55--60 & $60.0_{ -15.0 }^{ +15.0 }$ & $55.0_{ -15.0 }^{ +20.0 }$ & $57.5_{ -12.9 }^{ +17.5 }$ & $60.0_{ -20.0 }^{ +15.0 }$ \\
60--65 & $60.0_{ -13.2 }^{ +15.0 }$ & $60.0_{ -20.0 }^{ +15.0 }$ & $60.0_{ -11.0 }^{ +10.0 }$ & $60.0_{ -18.0 }^{ +10.0 }$ \\
65--70 & $60.0_{ -15.0 }^{ +15.0 }$ & $60.0_{ -15.0 }^{ +15.0 }$ & $65.0_{ -15.0 }^{ +10.0 }$ & $65.0_{ -15.8 }^{ +15.0 }$ \\
70--75 & $65.0_{ -11.4 }^{ +10.0 }$ & $65.0_{ -20.0 }^{ +10.0 }$ & $65.0_{ -10.8 }^{ +10.8 }$ & $65.0_{ -10.0 }^{ +10.0 }$ \\
75--80 & $70.0_{ -20.0 }^{ +10.0 }$ & $60.0_{ -10.2 }^{ +3.4 }$ & $70.0_{ -15.0 }^{ +10.0 }$ & $70.0_{ -14.0 }^{ +10.0 }$ \\
\enddata
\tablecomments{Simulations only go up to 80,000\,K. The true upper bound is likely much higher for measured temperatures above 40,000\,K (the true temperature could actually be unbounded from above in certain regimes; see the text for details).}
\end{deluxetable*}

Our results are truncated at 80,000\,K because those are the highest temperatures we simulated, but it is evident from the flattening of the curve in the top left panels of Figures \ref{fig:res_t_sdss_p05}, \ref{fig:res_t_sdss_p1}, \ref{fig:res_t_john_p05} and \ref{fig:res_t_john_p1}, that any optical measurement yielding a best-fit temperature of 60,000\,K or above for 0.05 magnitude uncertainties, and 40,000\,K or above for 0.1 magnitude uncertainties, actually sets only a lower limit on the true temperature. 

\subsection{Errors in Bolometric Luminosity Estimation}

In many cases, the blackbody temperature $T$ and radius $R$ are used to estimate the bolometric luminosity through $L_{bol}=4{\pi}R^2\sigma_{SB}T^4$ (with $\sigma_{SB}$ the Stephan-Boltzmann constant). One would expect that large errors in the temperature would translate to order-of-magnitude errors in the bolometric luminosity. However, as seen in the example fits in Figure \ref{fig:eg_fits}, there is typically an anticorrelated degeneracy between the temperature and radius, which cancels some of the error in each parameter when estimating the bolometric luminosity.

\begin{figure*}
\includegraphics[width=0.5\textwidth]{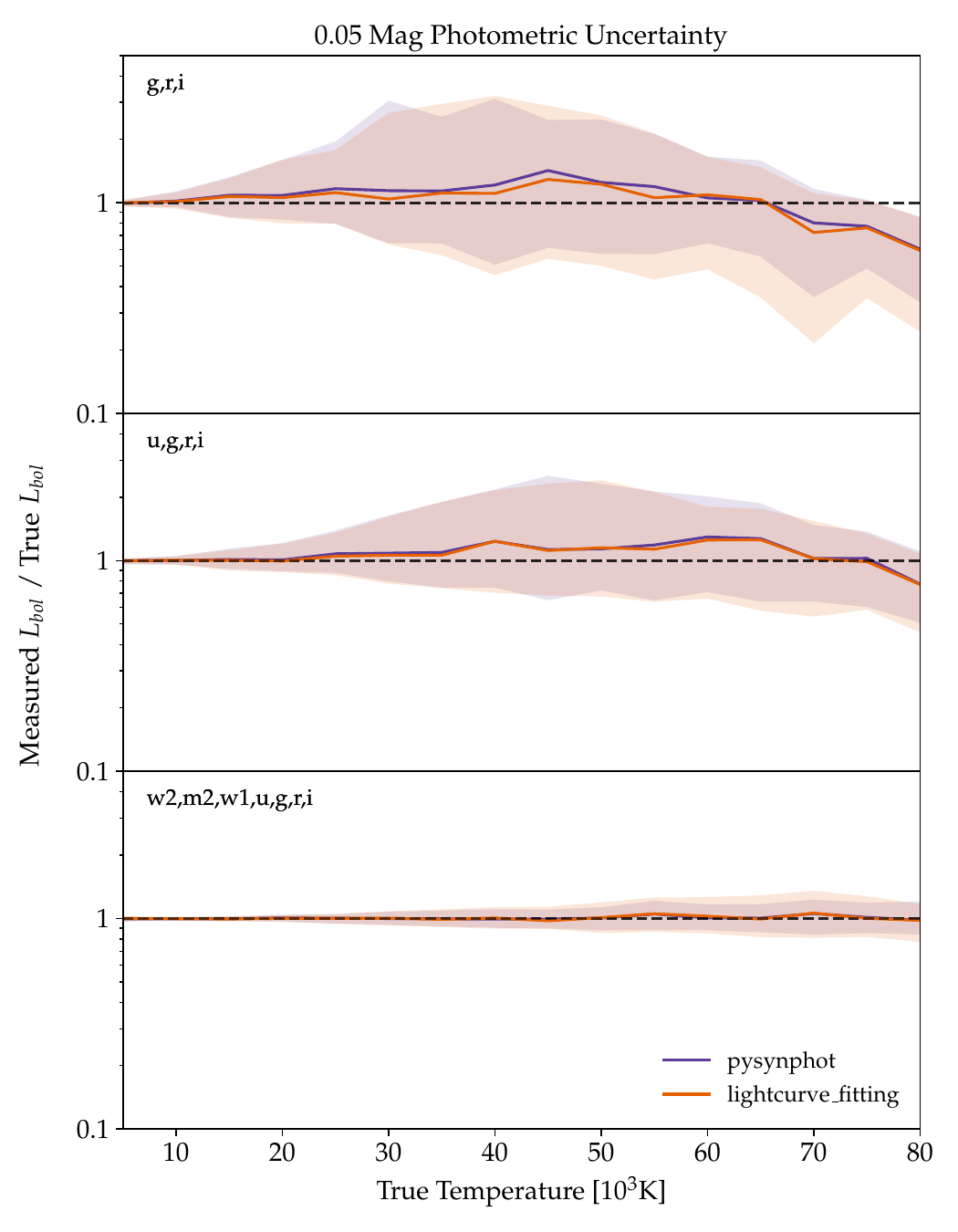}
\includegraphics[width=0.5\textwidth]{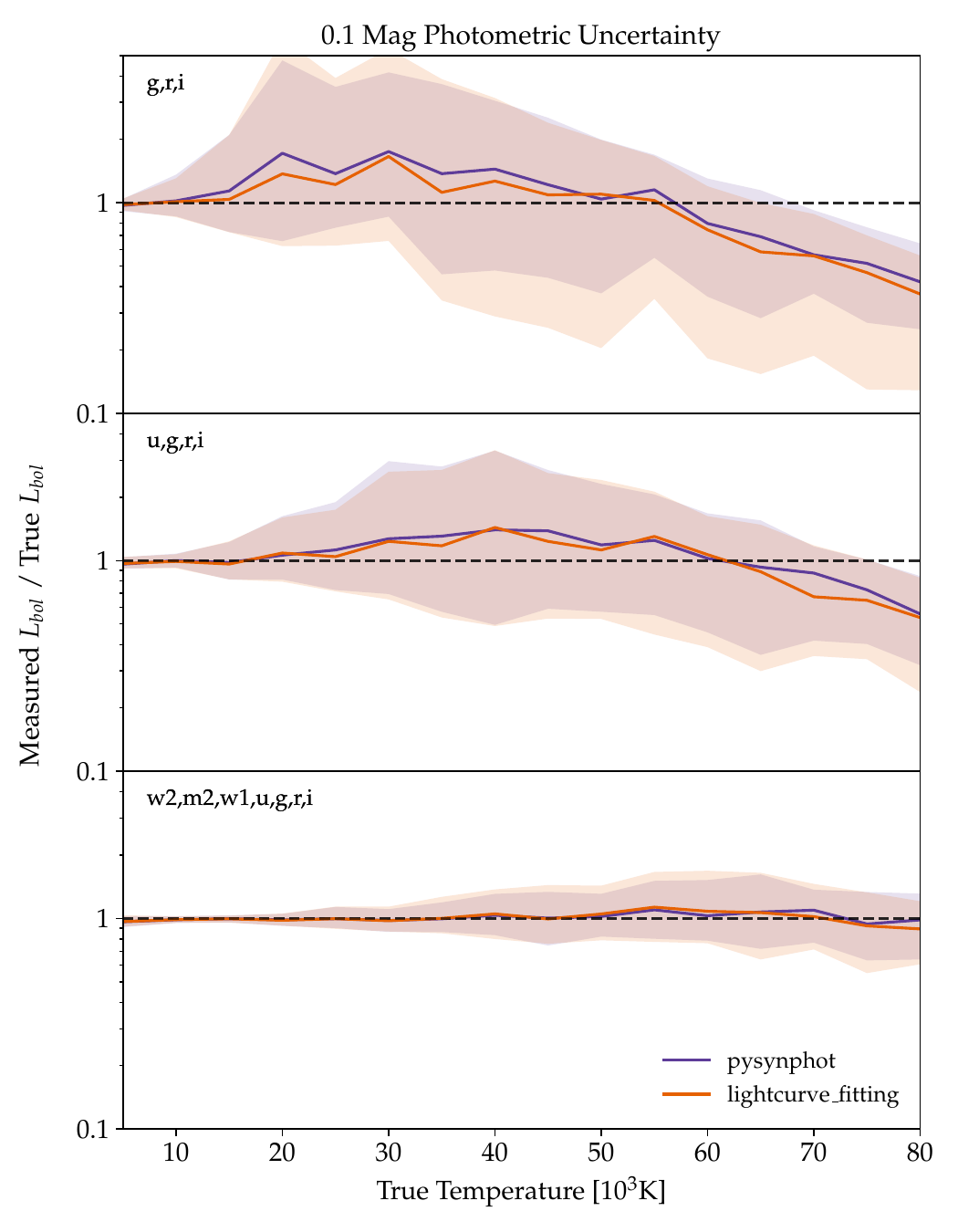}
\caption{\label{fig:res_l_sdss}Ratios between the measured and true bolometric luminosities at various temperatures for different SEDs as noted, using photometric uncertainties of 0.05 magnitudes (left) and 0.1 magnitudes (right). The solid lines use the bolometric luminosity calculated from the ensemble median best-fit temperature and radius, and the shaded regions denote the ensemble 16th--84th percentile range.} 
\end{figure*}

Figure \ref{fig:res_l_sdss} shows the ratio between the measured bolometric luminosity and the true bolometric luminosity for various SDSS and ultraviolet band combinations. When relying on optical data alone, the luminosities of blackbodies up to $\sim$60,000\,K are overestimated on average. For 0.05 magnitude uncertainties, the luminosity overestimation reaches a factor of $\sim$3 (84th ensemble percentile). At higher temperatures, beyond $\sim$60,000\,K, the preference shifts to underestimating the luminosity by factors of $\sim$3 (16th percentile). For 0.1 magnitude uncertainties, the luminosity over- and underestimates reach factors of $\sim$4--5 (at the 84th and 16th percentiles, respectively). 
Adding space-based ultraviolet bands reduces these errors significantly with the ensemble average luminosities tracking the true luminosity at all temperatures tested. Similar results are seen for the Johnson-Cousins SEDs (Fig. \ref{fig:res_l_john}). Like in the temperature estimation, there are almost no differences between the two methods (except at higher temperatures in the $gri$-band fits where the \texttt{lightcurve\_fitting} method has a longer tail toward underestimation of the true luminosity).

\section{Summary and Conclusions}

When limited to optical data, large errors in blackbody temperatures can occur, reaching $\sim$10,000\,K at 30,000--40,000\,K (depending on the measurement uncertainties). Beyond 40,000--60,000\,K it is only possible to obtain lower limits on the true temperature. It is important to consider these errors whenever an optical measurement yields a temperature $\gtrsim$ 25,000\,K. We calculate realistic uncertainties for such measurements and present them in Tables \ref{tab:inverse_t_flux} and \ref{tab:inverse_t_mag}. These results refer to fits using log-uniform priors. Uniform priors produce even less precise results, and are not discussed further. Underestimating the photometric uncertainties by a factor of two has almost no effect on the results (especially when using the \texttt{lightcurve\_fitting} package). 

Despite the large errors in temperature measurements of hot blackbodies with optical data alone, the error on the bolometric luminosity calculated from such fits is not orders of magnitude as one might naively expect from the $L_{bol}{\propto}T^4$ dependence. Temperature overestimations partially cancel with radius underestimations (and vice versa), producing bolometric luminosity errors up to factors of 3--5 in 68\% of cases. 

Adding ground-based ultraviolet observations reduces these errors somewhat, but space-based ultraviolet observations are required to reduce these errors significantly to less than a few hundred K in temperature and 10\%-20\% in bolometric luminosity.

There are no major differences between the results obtained when using the \texttt{pysynphot} method compared to the \texttt{lightcurve\_fitting} method. However, the \texttt{lightcurve\_fitting} method is significantly faster.

Our results relate to pure and single blackbodies. Departures from such spectra due to absorption and emission features, extinction, or other types of underlying spectra altogether, will induce even larger errors. In addition, errors in distance estimation will cause errors in determining the blackbody radius, which in turn will cause errors in determining the bolometric luminosity. Given the degeneracy between temperature and radius fitting, distance errors could cause additional errors also in temperature determination. 

Observations in the ultraviolet, currently possible in large numbers only with the {\it Swift} observatory, and soon also with the Ultraviolet Transient Astronomy Satellite \citep[ULTRASAT;][]{Sagiv2014}, are crucial for constraining the emission properties, and hence physics, of high-temperature astrophysical objects and events. When ultraviolet observations are not available, it is important to be aware of the uncertainties quantified here. 

~\\
We thank G. Hosseinzadeh for assistance in using the \texttt{lightcurve\_fitting} package and for helpful comments, and D. A. Howell for helpful comments.
I.A. is a CIFAR Azrieli Global Scholar in the Gravity and the Extreme Universe Program and acknowledges support from that program, from the European Research Council (ERC) under the European Union’s Horizon 2020 research and innovation program (grant agreement number 852097), from the Israel Science Foundation (grant number 2752/19), from the United States -- Israel Binational Science Foundation (BSF), and from the Israeli Council for Higher Education Alon Fellowship.

\software{PySynphot \citep{pysynphot}, 
Lightcurve\_Fitting \citep{lightcurve_fitting},
emcee \citep{Foreman-Mackey2013}.}

\bibliography{refs}{}

\begin{thebibliography}{}
\expandafter\ifx\csname natexlab\endcsname\relax\def\natexlab#1{#1}\fi
\providecommand{\url}[1]{\href{#1}{#1}}
\providecommand{\dodoi}[1]{doi:~\href{http://doi.org/#1}{\nolinkurl{#1}}}
\providecommand{\doeprint}[1]{\href{http://ascl.net/#1}{\nolinkurl{http://ascl.net/#1}}}
\providecommand{\doarXiv}[1]{\href{https://arxiv.org/abs/#1}{\nolinkurl{https://arxiv.org/abs/#1}}}

\bibitem[{{Abbott} {et~al.}(2017){Abbott}, {Abbott}, {Abbott}, \&
  et~al.}]{Abbott2017}
{Abbott}, B.~P., {Abbott}, R., {Abbott}, T.~D., \& et~al. 2017, \apjl, 848,
  L12, \dodoi{10.3847/2041-8213/aa91c9}

\bibitem[{{Arcavi}(2018)}]{Arcavi2018}
{Arcavi}, I. 2018, \apjl, 855, L23, \dodoi{10.3847/2041-8213/aab267}

\bibitem[{{Arcavi} {et~al.}(2017){Arcavi}, {Hosseinzadeh}, {Brown}, {Smartt},
  {Valenti}, {Tartaglia}, {Piro}, {Sanchez}, {Nicholls}, {Monard}, {Howell},
  {McCully}, {Sand}, {Tonry}, {Denneau}, {Stalder}, {Heinze}, {Rest}, {Smith},
  \& {Bishop}}]{Arcavi2017}
{Arcavi}, I., {Hosseinzadeh}, G., {Brown}, P.~J., {et~al.} 2017, \apjl, 837,
  L2, \dodoi{10.3847/2041-8213/aa5be1}

\bibitem[{{Breeveld} {et~al.}(2011){Breeveld}, {Landsman}, {Holland}, {Roming},
  {Kuin}, \& {Page}}]{Breeveld2011}
{Breeveld}, A.~A., {Landsman}, W., {Holland}, S.~T., {et~al.} 2011, in American
  Institute of Physics Conference Series, Vol. 1358, Gamma Ray Bursts 2010, ed.
  J.~E. {McEnery}, J.~L. {Racusin}, \& N.~{Gehrels}, 373--376,
  \dodoi{10.1063/1.3621807}

\bibitem[{{Brown} {et~al.}(2016){Brown}, {Breeveld}, {Roming}, \&
  {Siegel}}]{Brown2016}
{Brown}, P.~J., {Breeveld}, A., {Roming}, P. W.~A., \& {Siegel}, M. 2016, \aj,
  152, 102, \dodoi{10.3847/0004-6256/152/4/102}

\bibitem[{{Cousins}(1976)}]{Cousins1976}
{Cousins}, A.~W.~J. 1976, \memras, 81, 25

\bibitem[{{Dai} {et~al.}(2018){Dai}, {McKinney}, {Roth}, {Ramirez-Ruiz}, \&
  {Miller}}]{Dai2018}
{Dai}, L., {McKinney}, J.~C., {Roth}, N., {Ramirez-Ruiz}, E., \& {Miller},
  M.~C. 2018, \apjl, 859, L20, \dodoi{10.3847/2041-8213/aab429}

\bibitem[{{Doi} {et~al.}(2010){Doi}, {Tanaka}, {Fukugita}, {Gunn}, {Yasuda},
  {Ivezi{\'c}}, {Brinkmann}, {de Haars}, {Kleinman}, {Krzesinski}, \& {French
  Leger}}]{Doi2010}
{Doi}, M., {Tanaka}, M., {Fukugita}, M., {et~al.} 2010, \aj, 139, 1628,
  \dodoi{10.1088/0004-6256/139/4/1628}

\bibitem[{{Faran} {et~al.}(2018){Faran}, {Nakar}, \& {Poznanski}}]{Faran2018}
{Faran}, T., {Nakar}, E., \& {Poznanski}, D. 2018, \mnras, 473, 513,
  \dodoi{10.1093/mnras/stx2288}

\bibitem[{{Foreman-Mackey} {et~al.}(2013){Foreman-Mackey}, {Hogg}, {Lang}, \&
  {Goodman}}]{Foreman-Mackey2013}
{Foreman-Mackey}, D., {Hogg}, D.~W., {Lang}, D., \& {Goodman}, J. 2013, \pasp,
  125, 306, \dodoi{10.1086/670067}

\bibitem[{{Gehrels} {et~al.}(2004){Gehrels}, {Chincarini}, {Giommi}, {Mason},
  {Nousek}, {Wells}, {White}, {Barthelmy}, {Burrows}, {Cominsky}, {Hurley},
  {Marshall}, {M{\'e}sz{\'a}ros}, {Roming}, {Angelini}, {Barbier}, {Belloni},
  {Campana}, {Caraveo}, {Chester}, {Citterio}, {Cline}, {Cropper}, {Cummings},
  {Dean}, {Feigelson}, {Fenimore}, {Frail}, {Fruchter}, {Garmire}, {Gendreau},
  {Ghisellini}, {Greiner}, {Hill}, {Hunsberger}, {Krimm}, {Kulkarni}, {Kumar},
  {Lebrun}, {Lloyd-Ronning}, {Markwardt}, {Mattson}, {Mushotzky}, {Norris},
  {Osborne}, {Paczynski}, {Palmer}, {Park}, {Parsons}, {Paul}, {Rees},
  {Reynolds}, {Rhoads}, {Sasseen}, {Schaefer}, {Short}, {Smale}, {Smith},
  {Stella}, {Tagliaferri}, {Takahashi}, {Tashiro}, {Townsley}, {Tueller},
  {Turner}, {Vietri}, {Voges}, {Ward}, {Willingale}, {Zerbi}, \&
  {Zhang}}]{Gehrels2004}
{Gehrels}, N., {Chincarini}, G., {Giommi}, P., {et~al.} 2004, \apj, 611, 1005,
  \dodoi{10.1086/422091}

\bibitem[{{Gezari}(2021)}]{Gezari2021}
{Gezari}, S. 2021, \araa, 59, 21, \dodoi{10.1146/annurev-astro-111720-030029}

\bibitem[{Hosseinzadeh \& Gomez(2021)}]{lightcurve_fitting}
Hosseinzadeh, G., \& Gomez, S. 2021, Light Curve Fitting, v0.3.0,  Zenodo,
  \dodoi{10.5281/zenodo.5520603}

\bibitem[{{Johnson} \& {Morgan}(1953)}]{Johnson1953}
{Johnson}, H.~L., \& {Morgan}, W.~W. 1953, \apj, 117, 313,
  \dodoi{10.1086/145697}

\bibitem[{{Mockler} {et~al.}(2019){Mockler}, {Guillochon}, \&
  {Ramirez-Ruiz}}]{Mockler2019}
{Mockler}, B., {Guillochon}, J., \& {Ramirez-Ruiz}, E. 2019, \apj, 872, 151,
  \dodoi{10.3847/1538-4357/ab010f}

\bibitem[{{Nakar} \& {Sari}(2010)}]{Nakar2010}
{Nakar}, E., \& {Sari}, R. 2010, \apj, 725, 904,
  \dodoi{10.1088/0004-637X/725/1/904}

\bibitem[{{Piran} {et~al.}(2015){Piran}, {Svirski}, {Krolik}, {Cheng}, \&
  {Shiokawa}}]{Piran2015}
{Piran}, T., {Svirski}, G., {Krolik}, J., {Cheng}, R.~M., \& {Shiokawa}, H.
  2015, \apj, 806, 164, \dodoi{10.1088/0004-637X/806/2/164}

\bibitem[{{Rabinak} \& {Waxman}(2011)}]{Rabinak2011}
{Rabinak}, I., \& {Waxman}, E. 2011, \apj, 728, 63,
  \dodoi{10.1088/0004-637X/728/1/63}

\bibitem[{{Roming} {et~al.}(2005){Roming}, {Kennedy}, {Mason}, {Nousek}, {Ahr},
  {Bingham}, {Broos}, {Carter}, {Hancock}, {Huckle}, {Hunsberger}, {Kawakami},
  {Killough}, {Koch}, {McLelland}, {Smith}, {Smith}, {Soto}, {Boyd},
  {Breeveld}, {Holland}, {Ivanushkina}, {Pryzby}, {Still}, \&
  {Stock}}]{Roming2005}
{Roming}, P. W.~A., {Kennedy}, T.~E., {Mason}, K.~O., {et~al.} 2005, \ssr, 120,
  95, \dodoi{10.1007/s11214-005-5095-4}

\bibitem[{{Rubin} \& {Gal-Yam}(2017)}]{Rubin2017}
{Rubin}, A., \& {Gal-Yam}, A. 2017, \apj, 848, 8,
  \dodoi{10.3847/1538-4357/aa8465}

\bibitem[{{Ryu} {et~al.}(2020){Ryu}, {Krolik}, \& {Piran}}]{Ryu2020}
{Ryu}, T., {Krolik}, J., \& {Piran}, T. 2020, \apj, 904, 73,
  \dodoi{10.3847/1538-4357/abbf4d}

\bibitem[{{Sagiv} {et~al.}(2014){Sagiv}, {Gal-Yam}, {Ofek}, {Waxman},
  {Aharonson}, {Kulkarni}, {Nakar}, {Maoz}, {Trakhtenbrot}, {Phinney}, {Topaz},
  {Beichman}, {Murthy}, \& {Worden}}]{Sagiv2014}
{Sagiv}, I., {Gal-Yam}, A., {Ofek}, E.~O., {et~al.} 2014, \aj, 147, 79,
  \dodoi{10.1088/0004-6256/147/4/79}

\bibitem[{{Sapir} \& {Waxman}(2017)}]{Sapir2017}
{Sapir}, N., \& {Waxman}, E. 2017, \apj, 838, 130,
  \dodoi{10.3847/1538-4357/aa64df}

\bibitem[{{Shussman} {et~al.}(2016){Shussman}, {Waldman}, \&
  {Nakar}}]{Shussman2016}
{Shussman}, T., {Waldman}, R., \& {Nakar}, E. 2016, arXiv e-prints,
  arXiv:1610.05323.
\newblock \doarXiv{1610.05323}

\bibitem[{{STScI Development Team}(2013)}]{pysynphot}
{STScI Development Team}. 2013, {pysynphot: Synthetic photometry software
  package}.
\newblock \doeprint{1303.023}

\bibitem[{{Valenti} {et~al.}(2016){Valenti}, {Howell}, {Stritzinger}, {Graham},
  {Hosseinzadeh}, {Arcavi}, {Bildsten}, {Jerkstrand}, {McCully}, {Pastorello},
  {Piro}, {Sand}, {Smartt}, {Terreran}, {Baltay}, {Benetti}, {Brown},
  {Filippenko}, {Fraser}, {Rabinowitz}, {Sullivan}, \& {Yuan}}]{Valenti2016}
{Valenti}, S., {Howell}, D.~A., {Stritzinger}, M.~D., {et~al.} 2016, \mnras,
  459, 3939, \dodoi{10.1093/mnras/stw870}

\bibitem[{{van Velzen} {et~al.}(2020){van Velzen}, {Holoien}, {Onori}, {Hung},
  \& {Arcavi}}]{vanVelzen2020}
{van Velzen}, S., {Holoien}, T. W.~S., {Onori}, F., {Hung}, T., \& {Arcavi}, I.
  2020, \ssr, 216, 124, \dodoi{10.1007/s11214-020-00753-z}

\end{thebibliography}
\bibliographystyle{aasjournal}

\restartappendixnumbering

\appendix

\section{Differences in Synthetic Magnitudes Generated by the Two Packages} \label{sec:app-diffs}

\begin{figure}[b]
\includegraphics[width=\columnwidth]{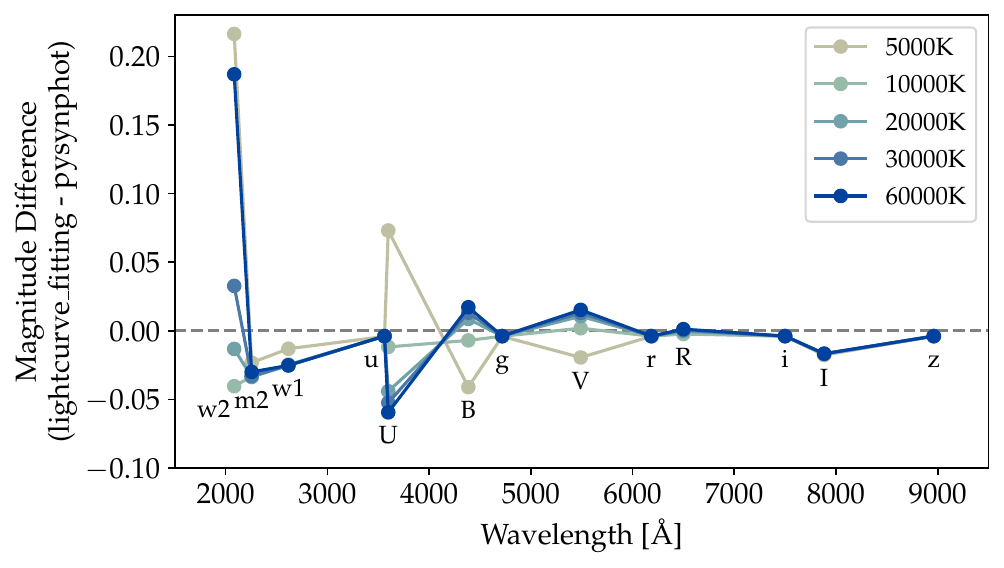}
\caption{\label{fig:methodsdiff}Differences in blackbody absolute magnitudes generated by the two packages used here (\texttt{pysynphot} and \texttt{lightcurve\_fitting}) for various bands at various temperatures.} 
\end{figure}

We generate synthetic magnitudes for blackbodies with a radius of 1,000\,R$_{\sun}$ and temperatures of 5,000, 10,000, 20,000, 30,000 and 60,000\,K using both the \texttt{pysynphot} and \texttt{lightcurve\_fitting} packages. Figure \ref{fig:methodsdiff} shows the differences in magnitudes generated by the two methods.

Some of these difference may be due to different filter response curves used by each package. However, the largest differences (up to 0.2 magnitudes for some temperatures) are seen in the $w2$ band even though both packages use the same \cite{Breeveld2011} response curves for the {\it Swift} $w1$, $m2$ and $w2$ bands. Therefore there are likely other factors responsible for these inconsistencies.

\section{Results and Analysis for Additional Scenarios}

\begin{figure*}
\includegraphics[width=\textwidth]{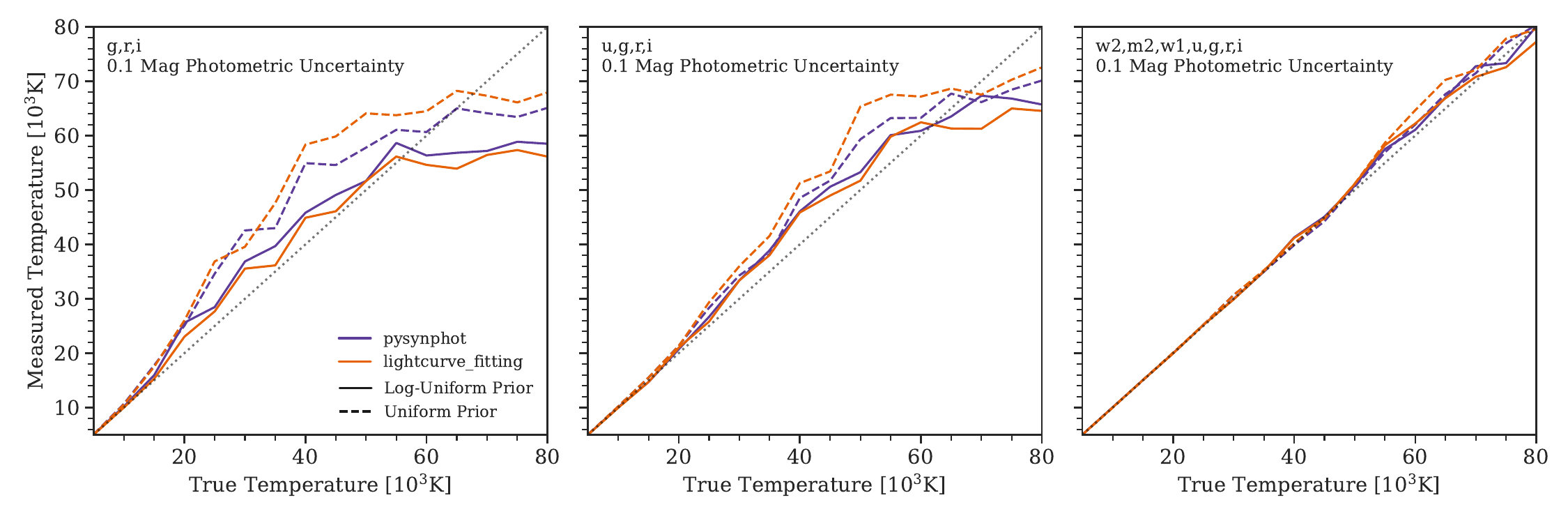}
\caption{\label{fig:res_t_sdss_p1_priorcomp}Same as Figure \ref{fig:res_t_sdss_p05_priorcomp} but using simulated measurement uncertainties of 0.1 magnitudes.}
\end{figure*}

\begin{figure*}
\includegraphics[width=\textwidth]{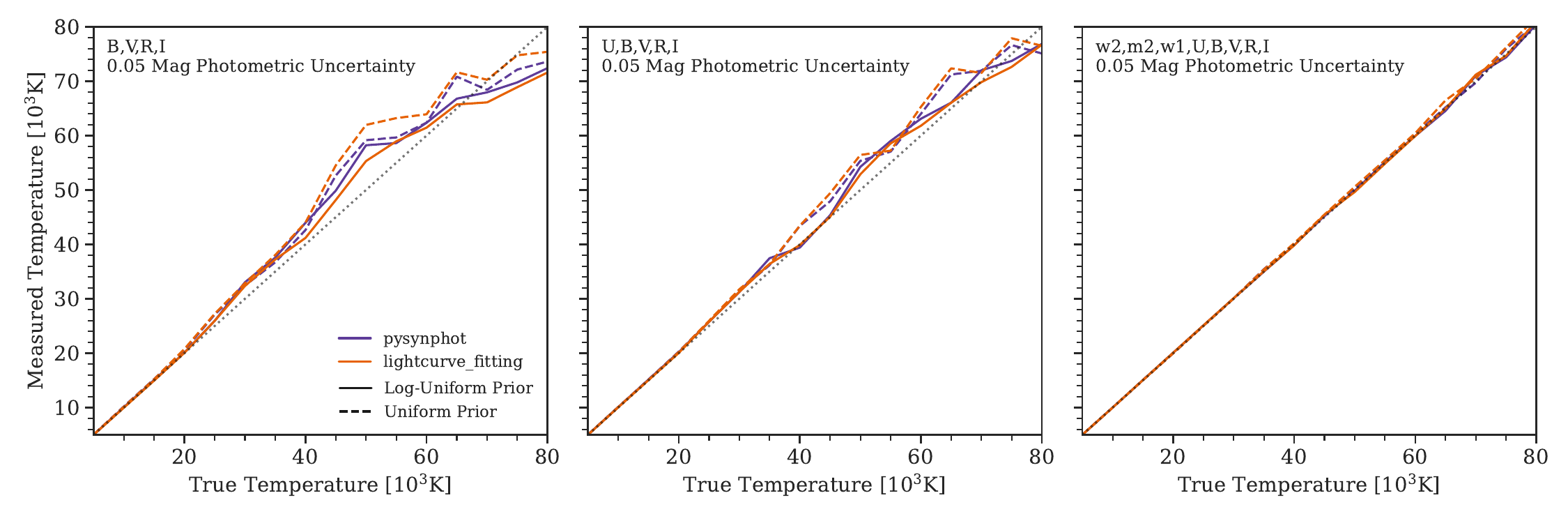}
\caption{\label{fig:res_t_john_p05_priorcomp}Same as Figure \ref{fig:res_t_sdss_p05_priorcomp} but for Johnson-Cousins bands.}
\end{figure*}

\begin{figure*}
\includegraphics[width=\textwidth]{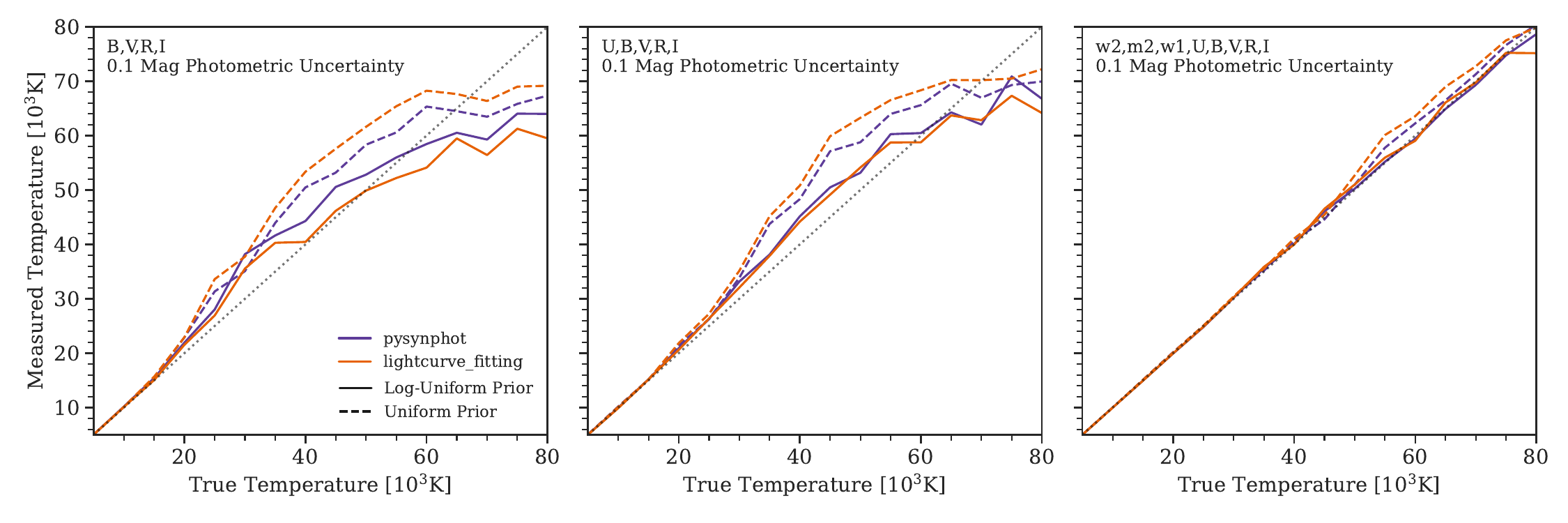}
\caption{\label{fig:res_t_john_p1_priorcomp}Same as Figure \ref{fig:res_t_sdss_p05_priorcomp} but for Johnson-Cousins bands and using simulated measurement uncertainties of 0.1 magnitudes.}
\end{figure*}

We present in Figures \ref{fig:res_t_sdss_p1_priorcomp}, \ref{fig:res_t_john_p05_priorcomp} and \ref{fig:res_t_john_p1_priorcomp} our prior comparison results for the SDSS SEDs with 0.1 magnitude errors and the Johnson-Cousins SEDs with both the 0.05 and 0.1 magnitude errors. The results are similar to those presented and analysed in the main text for the SDSS SEDs with 0.05 magnitude errors. 

\begin{figure*}
\includegraphics[width=\textwidth]{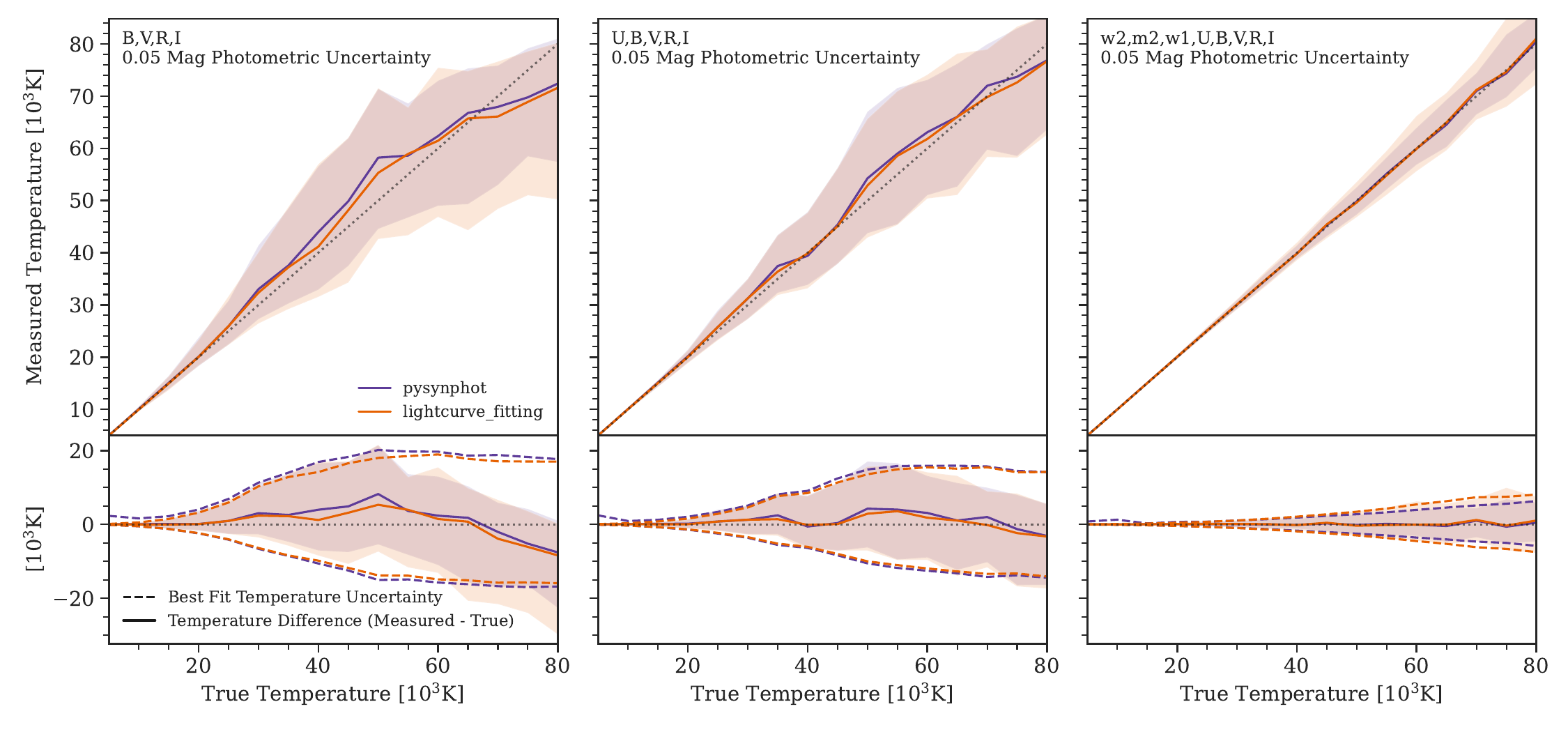}
\caption{\label{fig:res_t_john_p05}Same as Figure \ref{fig:res_t_sdss_p05} but for Johnson-Cousins bands.}
\end{figure*}

\begin{figure*}
\includegraphics[width=\textwidth]{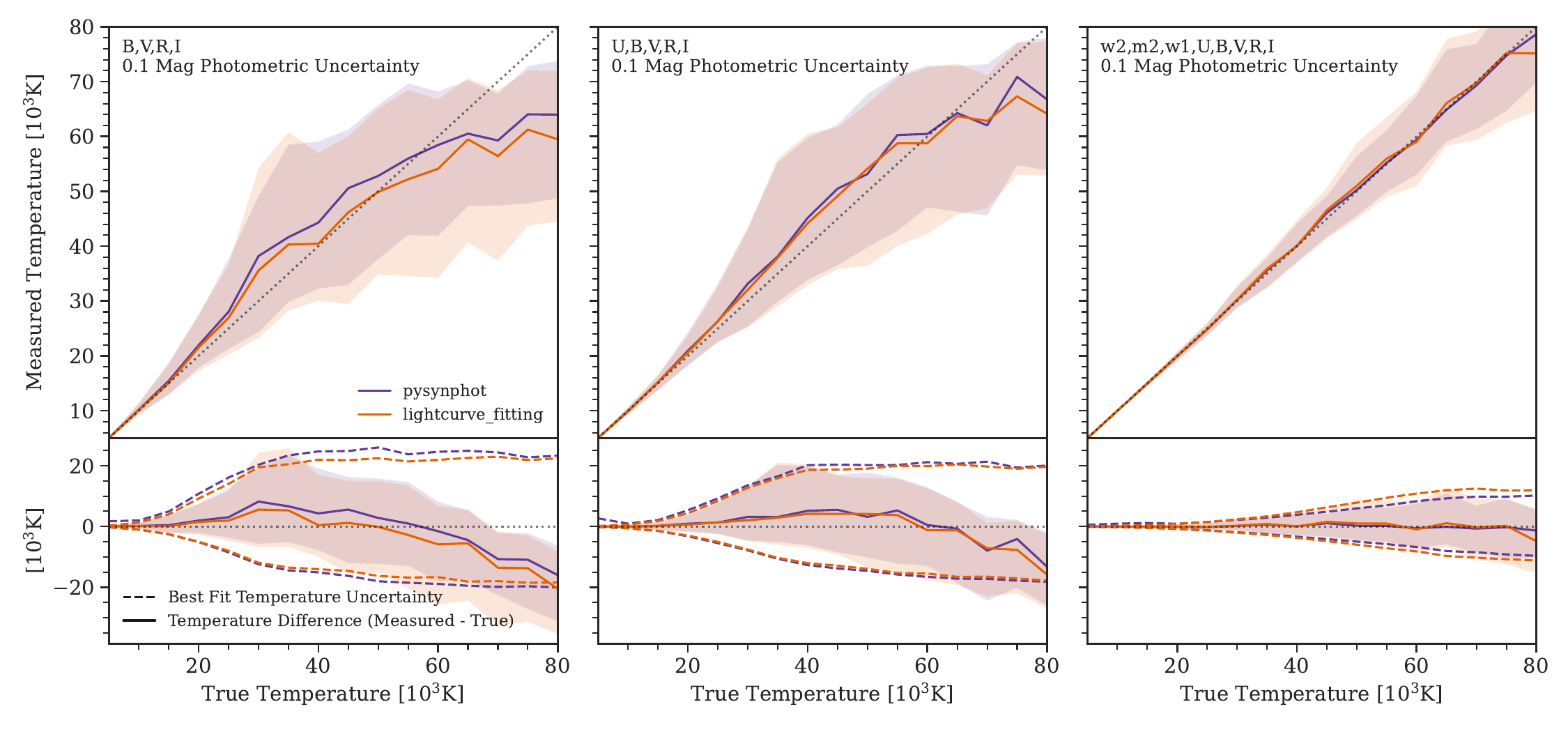}
\caption{\label{fig:res_t_john_p1}Same as Figure \ref{fig:res_t_sdss_p05} but for Johnson-Cousins bands and using simulated measurement uncertainties of 0.1 magnitudes.}
\end{figure*}

We present in Figures \ref{fig:res_t_john_p05} and \ref{fig:res_t_john_p1} our full simulation results for the Johnson-Cousins SEDs using the log-uniform priors. The results are similar to those presented and analysed in the main text for the SDSS SEDs with the same priors. 

\begin{figure*}
\includegraphics[width=\textwidth]{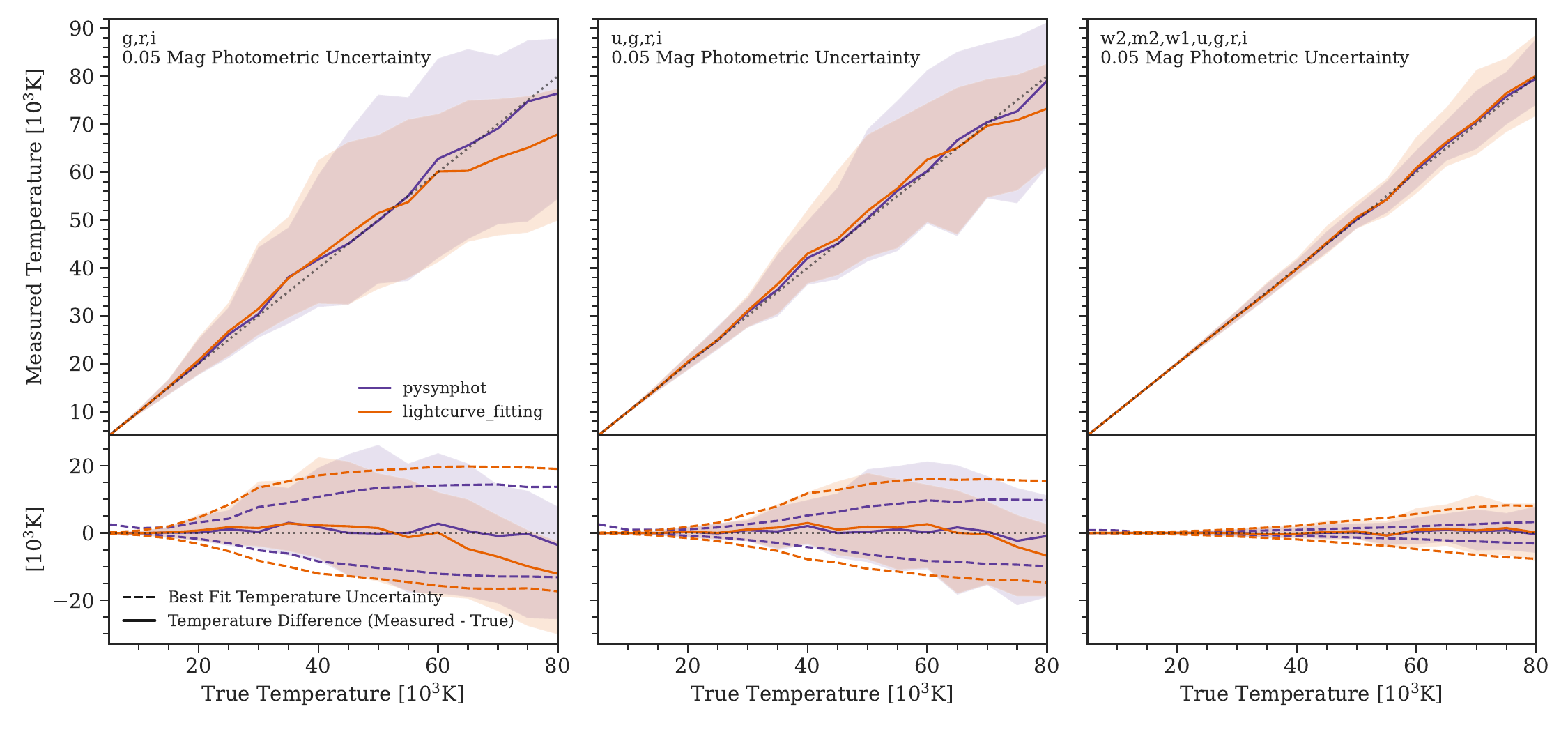}
\caption{\label{fig:res_t_sdss_p05_err_und}Same as Figure \ref{fig:res_t_sdss_p05} but for fits with uncertainties underestimated by a factor of two.}
\end{figure*}

\begin{figure*}
\includegraphics[width=\textwidth]{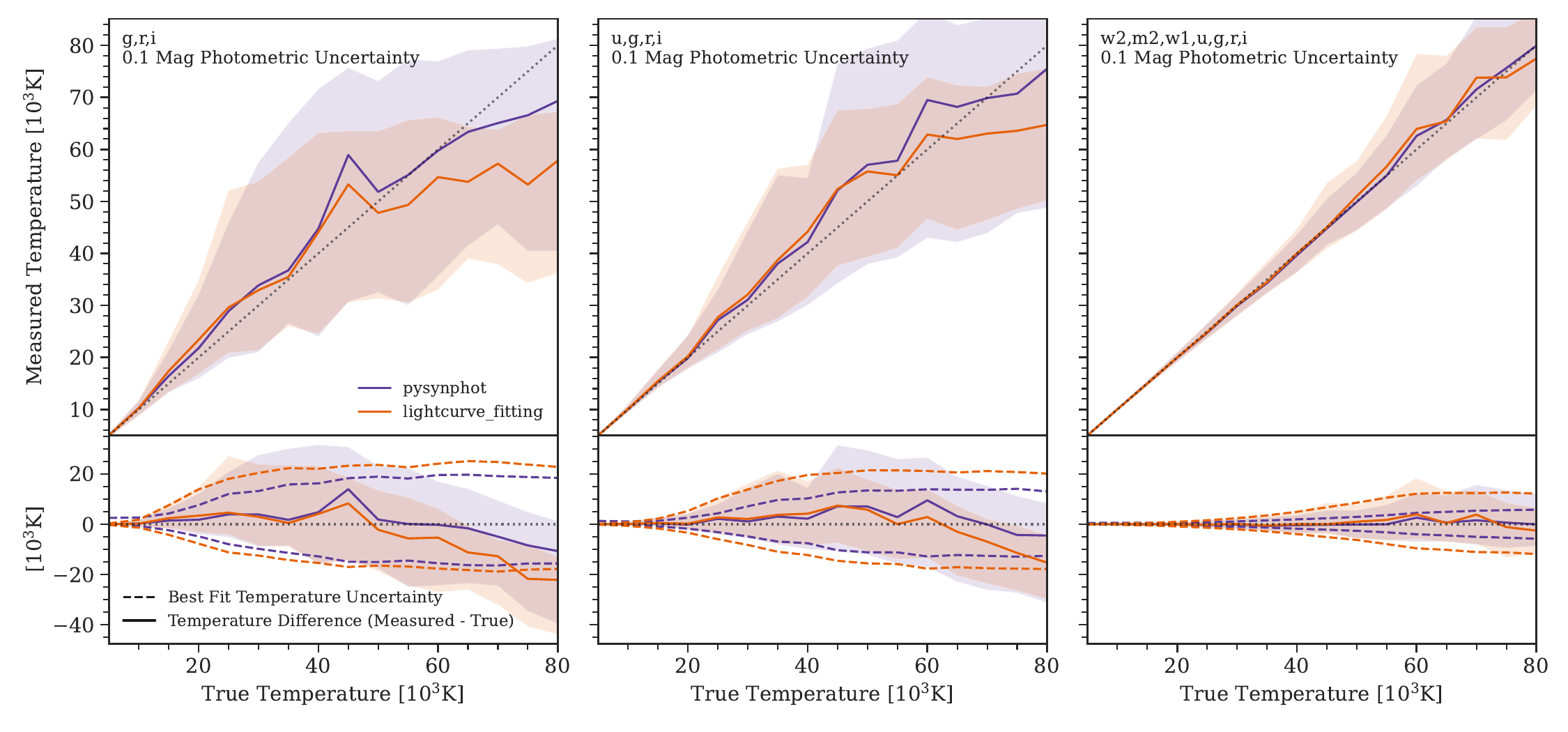}
\caption{\label{fig:res_t_sdss_p1_err_und}Same as Figure \ref{fig:res_t_sdss_p05_err_und} but using simulated measurement uncertainties of 0.1 magnitudes.}
\end{figure*}

\begin{figure*}
\includegraphics[width=\textwidth]{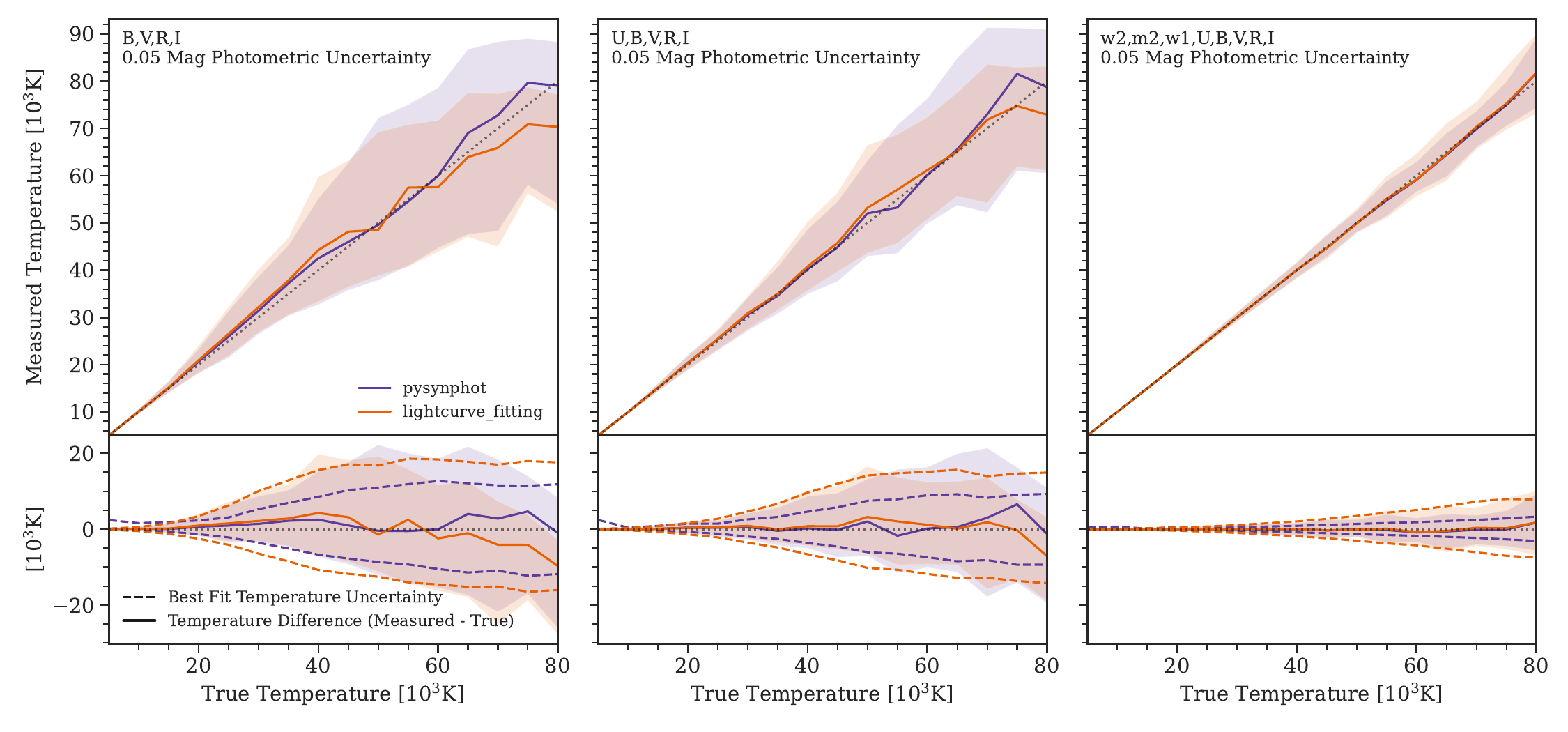}
\caption{\label{fig:res_t_john_p05_err_und}Same as Figure \ref{fig:res_t_sdss_p05_err_und} but for Johnson-Cousins bands.}
\end{figure*}

\begin{figure*}
\includegraphics[width=\textwidth]{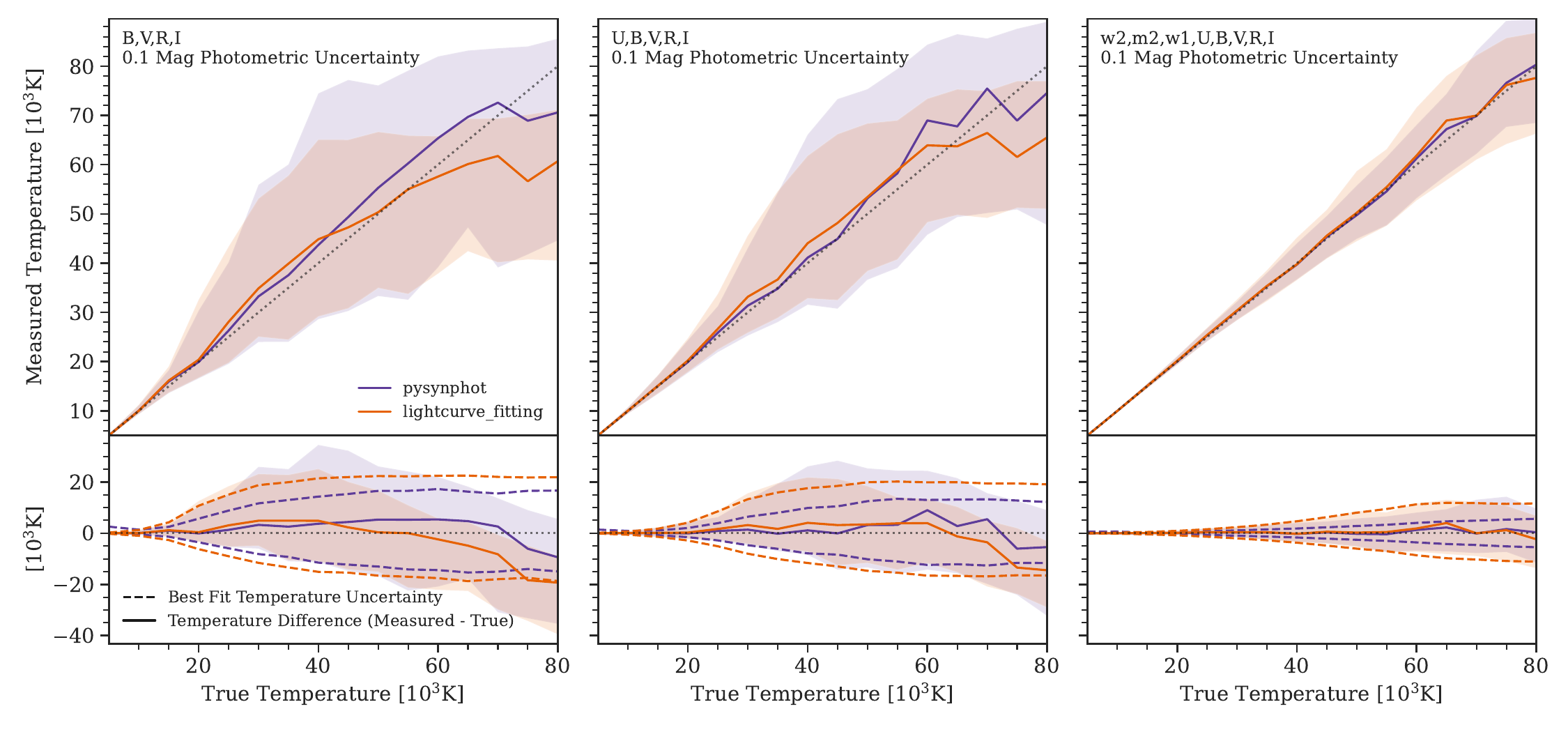}
\caption{\label{fig:res_t_john_p1_err_und}Same as Figure \ref{fig:res_t_sdss_p05_err_und} but for Johnson-Cousins bands and using simulated measurement uncertainties of 0.1 magnitudes.}
\end{figure*}

In Figures \ref{fig:res_t_sdss_p05_err_und} -- \ref{fig:res_t_john_p1_err_und} we present the full simulation results for the fits with uncertainties underestimated by a factor of two. We find no major differences compared to the fits with correctly estimated uncertainties, except that the uncertainties in the fitted temperature are underestimated when using the \texttt{pysynphot} method.

\begin{figure*}
\includegraphics[width=\textwidth]{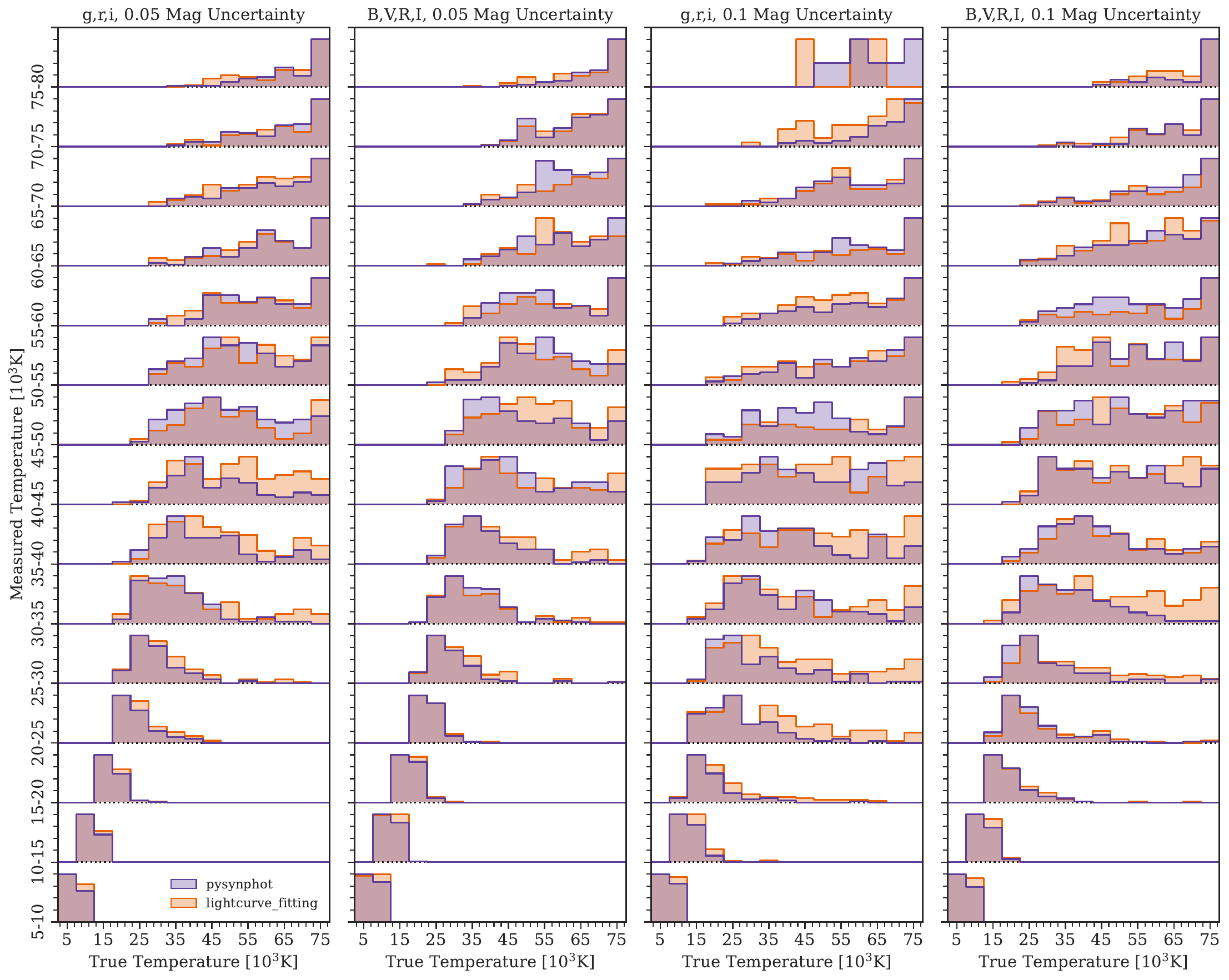}
\caption{\label{fig:inverse_t_all}Same as Figure \ref{fig:inverse_t_eg} but for both methods as shown in the legend, and for both SDSS and Johnson-Cousins optical bands, and various magnitude uncertainties, as shown in the titles.} 
\end{figure*}

The posterior distributions of true temperatures that produce various measured temperatures for the different optical SEDs, assumed magnitude uncertainties, and methods (using log-uniform priors) are shown in Figure \ref{fig:inverse_t_all}. The median and $1\sigma$ bounds of these distributions for the \texttt{pysynphot} method are presented in Table \ref{tab:inverse_t_mag}.

\begin{deluxetable*}{@{\extracolsep{4pt}}ccccc@{}}
\tablecaption{Same as Table \ref{tab:inverse_t_flux} but for the \texttt{pysynphot} method.} \label{tab:inverse_t_mag}
\tablehead{
\colhead{Measured temperature} & \multicolumn{4}{c}{True temperature} \\
\cline{1-1} \cline{2-5}
\colhead{} & \multicolumn{2}{c}{For a $gri$-band measurement} & \multicolumn{2}{c}{For a $BVRI$-band measurement}\\
\cline{2-3} \cline{4-5}
\colhead{} & \colhead{0.05 mag uncertainty} & \colhead{0.1 mag uncertainty} & \colhead{0.05 mag uncertainty} & \colhead{0.1 mag uncertainty} \\
\colhead{($10^3$\,K)} & \colhead{($10^3$\,K)} & \colhead{($10^3$\,K)} & \colhead{($10^3$\,K)} & \colhead{($10^3$\,K)}}
\startdata
5--10 & $5.0_{ -0.0 }^{ +5.0 }$ & $5.0_{ -0.0 }^{ +5.0 }$ & $5.0_{ -0.0 }^{ +5.0 }$ & $5.0_{ -0.0 }^{ +5.0 }$ \\
10--15 & $10.0_{ -0.0 }^{ +5.0 }$ & $10.0_{ -0.0 }^{ +5.0 }$ & $10.0_{ -0.0 }^{ +5.0 }$ & $10.0_{ -0.0 }^{ +5.0 }$ \\
15--20 & $15.0_{ -0.0 }^{ +5.0 }$ & $15.0_{ -0.0 }^{ +10.0 }$ & $15.0_{ -0.0 }^{ +5.0 }$ & $20.0_{ -5.0 }^{ +5.0 }$ \\
20--25 & $25.0_{ -5.0 }^{ +5.0 }$ & $25.0_{ -8.6 }^{ +10.0 }$ & $22.5_{ -2.5 }^{ +2.5 }$ & $25.0_{ -5.0 }^{ +10.0 }$ \\
25--30 & $30.0_{ -5.0 }^{ +5.0 }$ & $30.0_{ -10.0 }^{ +15.0 }$ & $30.0_{ -5.0 }^{ +5.0 }$ & $25.0_{ -5.0 }^{ +15.0 }$ \\
30--35 & $35.0_{ -10.0 }^{ +10.0 }$ & $35.0_{ -10.0 }^{ +20.0 }$ & $35.0_{ -6.0 }^{ +6.0 }$ & $35.0_{ -10.0 }^{ +15.0 }$ \\
35--40 & $40.0_{ -10.0 }^{ +15.0 }$ & $40.0_{ -15.0 }^{ +25.0 }$ & $35.0_{ -5.0 }^{ +15.0 }$ & $40.0_{ -10.0 }^{ +20.0 }$ \\
40--45 & $45.0_{ -10.0 }^{ +15.0 }$ & $45.0_{ -15.0 }^{ +20.0 }$ & $45.0_{ -10.0 }^{ +20.0 }$ & $50.0_{ -20.0 }^{ +17.0 }$ \\
45--50 & $50.0_{ -15.0 }^{ +18.4 }$ & $50.0_{ -20.0 }^{ +20.4 }$ & $45.0_{ -10.0 }^{ +20.0 }$ & $50.0_{ -15.0 }^{ +20.0 }$ \\
50--55 & $55.0_{ -15.0 }^{ +15.0 }$ & $60.0_{ -20.0 }^{ +15.0 }$ & $55.0_{ -10.0 }^{ +14.8 }$ & $55.0_{ -10.0 }^{ +16.6 }$ \\
55--60 & $60.0_{ -15.0 }^{ +15.0 }$ & $60.0_{ -20.0 }^{ +15.0 }$ & $55.0_{ -10.0 }^{ +20.0 }$ & $55.0_{ -15.0 }^{ +20.0 }$ \\
60--65 & $60.0_{ -15.0 }^{ +15.0 }$ & $60.0_{ -15.0 }^{ +15.0 }$ & $60.0_{ -10.0 }^{ +15.0 }$ & $60.0_{ -19.8 }^{ +15.0 }$ \\
65--70 & $65.0_{ -15.0 }^{ +10.0 }$ & $60.0_{ -15.0 }^{ +15.0 }$ & $62.5_{ -7.5 }^{ +12.5 }$ & $65.0_{ -15.0 }^{ +10.0 }$ \\
70--75 & $67.5_{ -17.5 }^{ +7.5 }$ & $70.0_{ -10.0 }^{ +10.0 }$ & $65.0_{ -15.0 }^{ +10.0 }$ & $65.0_{ -10.0 }^{ +15.0 }$ \\
75--80 & $70.0_{ -12.0 }^{ +10.0 }$ & $62.5_{ -6.9 }^{ +11.9 }$ & $75.0_{ -10.0 }^{ +5.0 }$ & $75.0_{ -17.8 }^{ +5.0 }$ \\
\enddata
\end{deluxetable*}

\begin{figure*}
\includegraphics[width=\columnwidth]{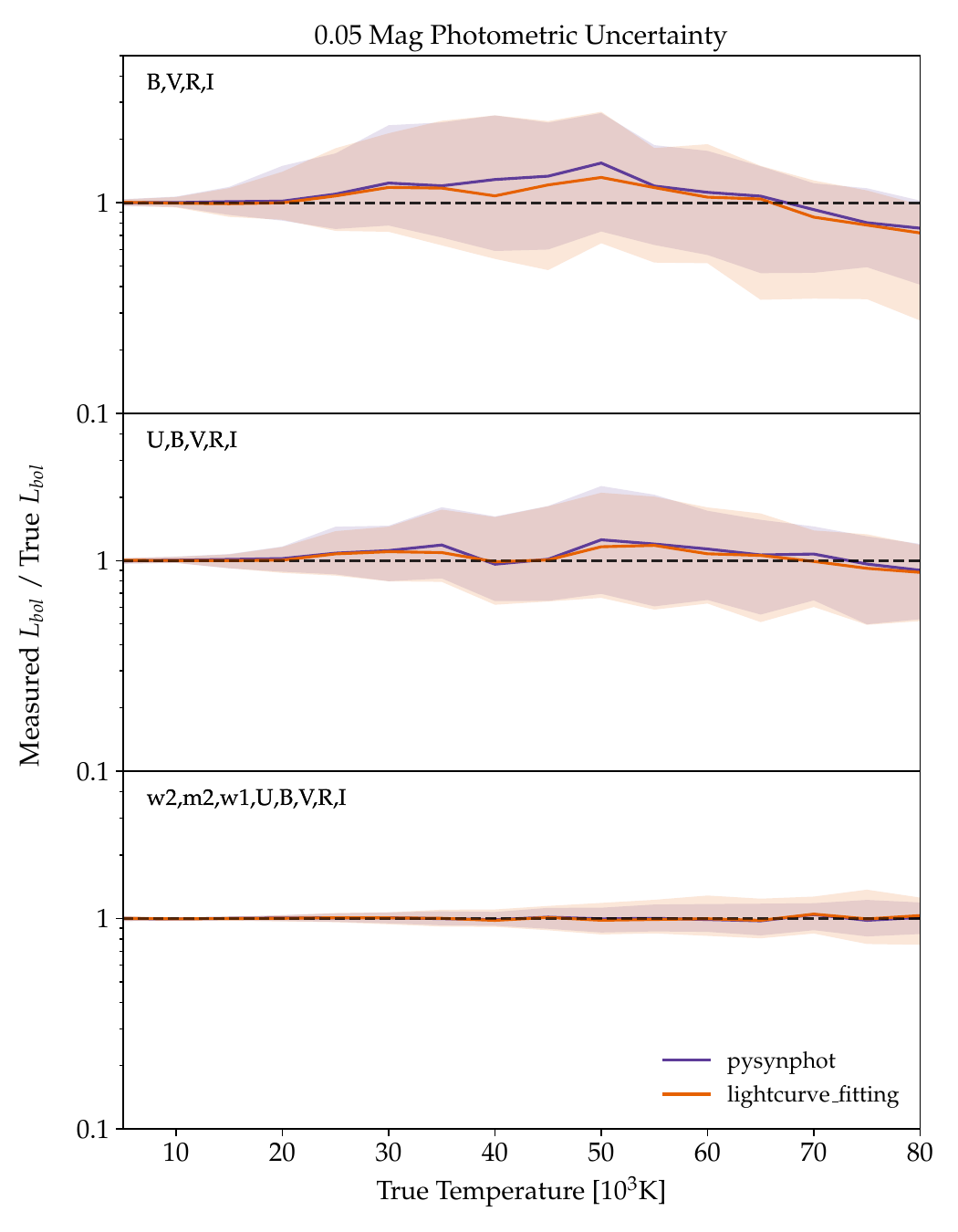}
\includegraphics[width=\columnwidth]{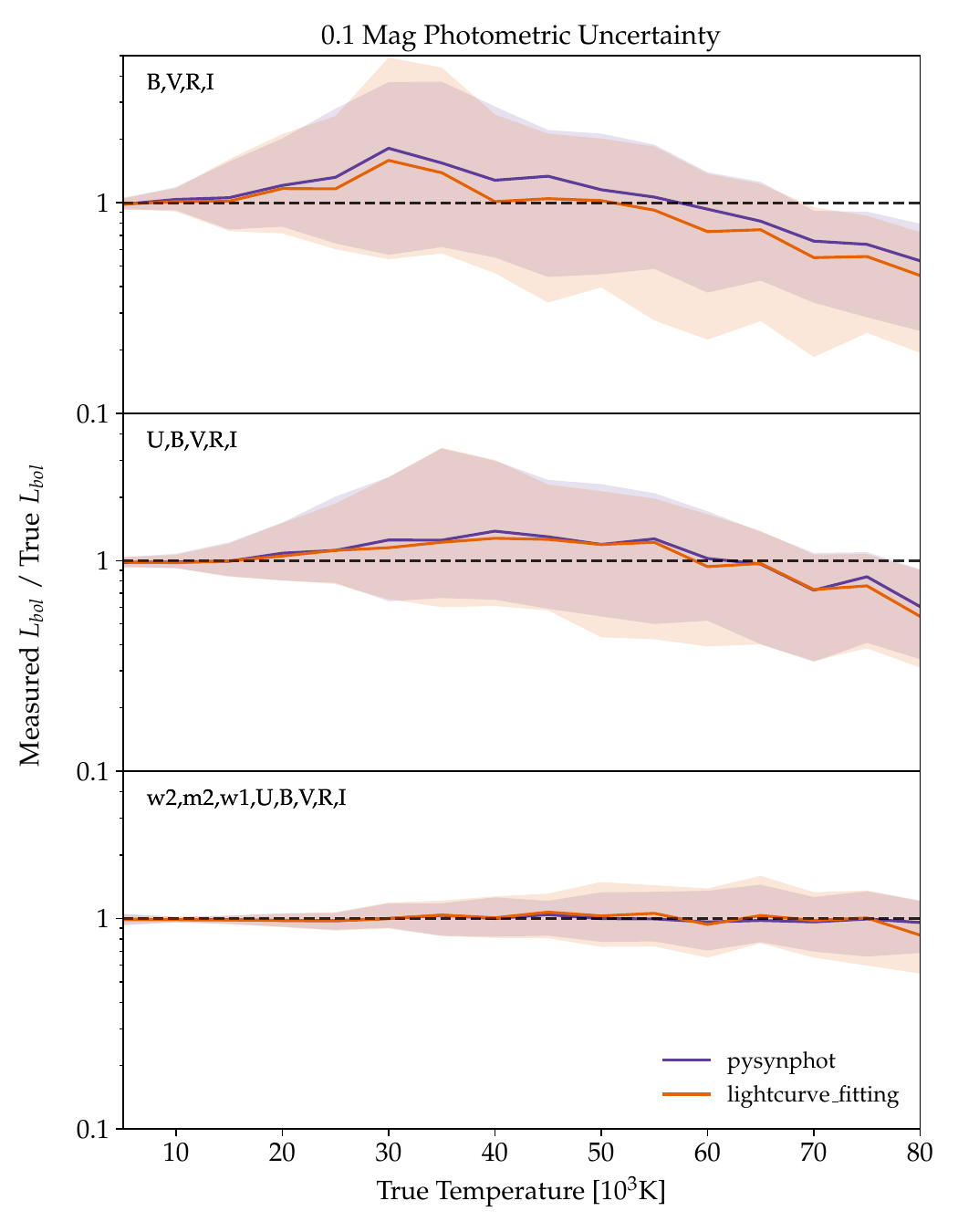}
\caption{\label{fig:res_l_john}Same as Figure \ref{fig:res_l_sdss} but but for Johnson-Cousins bands using 0.05 magnitude (left) and 0.1 magnitude (right) simulated measurement uncertainties.} 
\end{figure*}

The bolometric luminosity errors for the Johnson-Cousins SEDs (assuming log-uniform priors) are presented in Figure \ref{fig:res_l_john}. These results are similar to the results presented in the main text for the SDSS SEDs using the same priors.

\end{document}